\documentclass[pdftex,twocolumn,epjc3]{svjour3}          

\RequirePackage[T1]{fontenc}

\smartqed  

\RequirePackage{graphicx}
\RequirePackage{float}
\RequirePackage{flushend}
\RequirePackage{amsfonts}
\RequirePackage{amssymb}
\RequirePackage[numbers,sort&compress]{natbib}
\RequirePackage[colorlinks,citecolor=blue,urlcolor=blue,linkcolor=blue]{hyperref}

\journalname{Eur. Phys. J. B}

\begin{document}

\title{\color{black} The critical temperature of the 2D-Ising model through Deep Learning Autoencoders}

\author{Constantia Alexandrou\thanksref{addr1,addr2} \and Andreas Athenodorou \thanksref{e1,addr3} \and Charalambos~Chrysostomou \thanksref{addr1} \and Srijit Paul \thanksref{e2,addr1,addr4}
}                    

\institute{Computation-based Science and Technology Research Center, The Cyprus              Institute, 20 Kavafi Str., 2121, Nicosia,
            Cyprus.\label{addr1}
            \and 
            University of Cyprus, 1 Panepistimiou Avenue, 2109, Nicosia, Cyprus.\label{addr2} 
            \and 
            Dipartimento di Fisica, Universit\`a di Pisa and INFN, Sezione di Pisa, Largo Pontecorvo 3, 56127 Pisa, Italy.\label{addr3}
            \and 
            Faculty of Mathematics and Natural Sciences, University of Wuppertal, Gaussstr. 20, D-42119 Wuppertal.\label{addr4}
            }
\thankstext{e1}{e-mail:  andreas.athinodorou@pi.infn.it}
\thankstext{e2}{e-mail:  s.paul@hpc-leap.eu}

\date{Received: 23.10.2019 / Accepted: 2.10.2020}

\maketitle

\begin{abstract}
We investigate deep learning autoencoders for the unsupervised recognition of phase transitions in physical systems formulated on a lattice. We focus our investigation on the 2-dimensional ferromagnetic Ising model and then test the application of the autoencoder on the anti-ferromagnetic Ising model. We use spin configurations produced for the 2-dimensional ferromagnetic and anti-ferromagnetic Ising model in zero external magnetic field. For the ferromagnetic Ising model, we study numerically the relation between one latent {\color{black} variable} extracted from the autoencoder to the critical temperature $T_c$. The proposed autoencoder reveals the two phases, one for which the spins are ordered and the other for which spins are disordered, reflecting the restoration of the $\mathbb{Z}_2$ symmetry as the temperature increases. We provide a finite volume analysis for a sequence of increasing lattice sizes. For the largest volume studied, the transition between the two phases occurs very close to the theoretically extracted critical temperature. We define as a quasi-order parameter the absolute average latent {\color{black} variable} ${\tilde z}$, which enables us to predict the critical temperature. One can define a latent susceptibility and use it to quantify the value of the critical temperature $T_c(L)$ at different lattice sizes and that these values suffer from only small finite scaling effects. We demonstrate that $T_c(L)$ extrapolates to the known theoretical value as $L \to \infty$ suggesting that the autoencoder can also be used to extract the critical temperature of the phase transition to an adequate precision. Subsequently, we test the application of the autoencoder on the anti-ferromagnetic Ising model, demonstrating that the proposed network can detect the phase transition successfully in a similar way.
\end{abstract}

\section{Introduction}
\label{sec:Introduction} 
Recent advances in the implementation of  Artificial Intelligence (AI) for physical systems, especially, on those which can be formulated on a lattice, appear to be suitable for observing the corresponding underlying phase structure~\cite{Carrasquilla2017,vanNieuwenburg2017,PhysRevB.94.195105,Broecker2017,ch2017machine,PhysRevE.96.022140,PhysRevE.95.062122,2017arXiv170700663B,PhysRevB.97.205110,PhysRevLett.120.257204,2018arXiv180402709Z,2018JChPh.149s4109J,PhysRevE.98.022138,2018arXiv180801731Z,Kashiwa:2018jdi,Giannetti:2018vif,Zhou:2018ill}. So far methods such as the Principal Component Analysis (PCA) \cite{PhysRevB.94.195105,vanNieuwenburg2017,PhysRevE.95.062122,wetzel2017unsupervised,Foreman:2018ktj}, Supervised Machine Learning (ML) \cite{Broecker2017,2018arXiv180402709Z,morningstar2017deep}, Restricted Boltzmann Machines (RBMs)~\cite{Cossu:2018pxj,Funai:2018esm}, as well as autoencoders~\cite{PhysRevE.95.062122,PhysRevE.96.022140} appear to successfully identify different phase regions of classical statistical systems, such as the 2-dimensional (2D) Ising model that describes the (anti)ferromagnetic-paramagnetic transition. These techniques were also applied on quantum statistical systems, such as the Hubbard model~\cite{ch2017machine} that describes the transition between conducting and insulating systems. Very recently, similar studies have been applied for simulations of quantum fields on the lattice, such as the $SU(2)$ gauge theory~\cite{Wetzel:2017ooo} with an increased  complexity in the data  due to the structure of the $SU(2)$ gauge group.

Trained neural networks can thus help distinguish phases in simple statistical systems -- the structure of which is known -- but, more importantly in more complex systems where the underlying phase structure is unknown. In this work, we would like to examine whether the proposed, fully-connected (Dense), deep learning autoencoder, which does not require supervision during training, can shed light on the phase structure of the 2D-Ising model {\color{black} and enable the identification of the critical temperature in the thermodynamic limit. Furthermore, we would like to investigate if deep learning autoencoders can be used as tools in order to extract physical observables with better statistical accuracy. This means to define new observables which demonstrate different features than the well-known quantities such as the order parameter of the theory.}

{\color{black} As mentioned above, this is not the only work where autoencoders are used. Autoencoders have been previously used for the identification of phase transitions in Refs.~\cite{PhysRevE.95.062122,PhysRevE.96.022140}. In Ref.~\cite{PhysRevE.95.062122} the authors studied the phase transition of the 2D-Ising model using in addition to PCA an autoencoder based on Convolutional Neural Networks (CNNs) and observed that they could actually "see" a transition and differentiate the two distinct states of the 2D-Ising model through the latent variable. Nevertheless, this result has been obtained for one lattice volume and a finite volume analysis\footnote{For reasons of clarity, we mention that finite volume analysis has been applied in the past on results extracted via supervised machine learning~\cite{zhang2019few} as well as from PCA analysis~\cite{li2019extracting}, however not from autoencoders.} has not been carried out. Thus, it is not clear whether the observed critical point converges to the known critical temperature in the limit of infinite volume.} 

{\color{black} In Ref.~\cite{PhysRevE.96.022140} the author used a variational autoencoder as well as an autoencoder to demonstrate that the latent parameters of the autoencoder, resulting by feeding it with configurations of the 2D-Ising model, corresponds to the known order parameter, i.e. the magnetization. Although this was carried out at one lattice volume, it is clear that since the latent variable is identified as the magnetization, the extracted critical temperature for finite volume will converge to the known, theoretically extracted, critical temperature at the infinite volume limit. At this point, it should be made clear that goal of this work is not to use the autoencoder to reproduce the order parameter, but as a technical tool enabling to study particular features of the model and identify new quantities which might be proven useful for analyzing the phase behaviour of statistical and possibly gauge theoretical models. It actually turns out that using the proposed network with the chosen activation functions, the latent variable we observe is not identified as the magnetization but as a different quantity which is affected less by finite volume effects, leading to faster convergence in the thermodynamic limit.}

{\color{black} Moreover, in the field of Computational Physics, autoencoders are also used by exploiting their generative context. In other words, there have been investigations on how to use variational autoencoders towards the reconstruction of physically meaningful configurations of statistical systems such as the 2D-Ising model~\cite{luchnikov2019variational} and the 2D $XY$ model~\cite{Cristoforetti:2017naf}. Although this is a hot and promising topic since it can potentially reduce the cost of the production of such configurations, the successful reconstruction of 2D-Ising configurations is beyond the scope of this work.}

Deep learning autoencoders are frequently used in cases where data hides interesting structure by processing the raw datasets. They can, therefore, be used to discover interesting structure in ensembles produced for a range of a parameter that characterizes the phase space of the model (with different sectors having different physical properties). One such example is the ferromagnetic Ising model for which at the critical temperature $T_c$, the system undergoes a transition from the ordered phase to the disordered. A minor variation of this model is the anti-ferromagnetic Ising model, where the system also undergoes a similar phase transition.

In this work, we investigate the action of unsupervised machine learning, namely the deep learning autoencoder (not variational), towards the identification of the phase transition of the 2D-Ising (anti)ferromagnetic model. More specifically, we produce decorrelated configurations for the 2D-Ising model for a given range of temperatures, and then we apply the autoencoder trying to understand what characteristics of the phase structure we can capture. Hence, technically, this work combines the production of configurations using Monte Carlo methods as well as the deep learning autoencoder algorithm. We observe that the autoencoder can capture the underlying $\mathbb{Z}_2$ symmetry and can indeed find out where the transition occurs by identifying a relevant, quasi-order parameter: the mean value of the absolute latent variable. Although this quantity is not suitable for predicting the order of the transition, it can determine the critical temperature with small finite scaling effects. 

This article is organized as follows: In section~\ref{sec:ising} we present a brief description of the 2D-Ising ferromagnetic and anti-ferromagnetic model, explaining the production of the configurations as well as its phase structure. In section~\ref{sec:autoencoders}, we discuss the deep learning autoencoder, explain how it works and provide the structure of the network. Subsequently, in section~\ref{sec:results} we provide our results for the ferromagnetic 2D-Ising model. Additionally, as a test, we apply our autoencoder on the anti-ferromagnetic 2D-Ising model and demonstrate our results in section~\ref{sec:results_antiferromagnetic}. Finally, in section~\ref{sec:conclusions}, we present our conclusions.

\section{\label{sec:ising} The 2-Dimensional Ising Model}

One of the most interesting physical phenomena in nature is magnetism. It is known that the ferromagnetic materials exhibit a spontaneous magnetization in the absence of an external magnetic field. Such magnetization occurs only if the temperature of the system is lower than a known critical temperature $T_c$, the so called Curie temperature. If the temperature of the system is raised so that  $T>T_c$, then the magnetization vanishes. In principle, the critical temperature $T_c$ separates the microstates of the system from being ordered or magnetized for $T<T_c$ to being randomly oriented resulting in zero magnetization; these two phases correspond to the ferromagnetic and the disordered phases, respectively.

(Anti)ferro-magnetism has a quantum mechanical nature and, thus, much effort is invested towards its understanding. Albeit quantum mechanical, simple classical models can help to gain insight into this effect. The 2D-Ising model is a classical model that is commonly used to study magnetism. The 2D-Ising model can be considered as a lattice with $N = N_x \times N_y$ sites, on each of which a double valued spin $s_i$ is located, either being in an "up" orientation denoted by $\uparrow$ or $s_i=+$ or "down" denoted by $\downarrow$ or $s_i=-$.  

The macroscopic properties of the 2D-Ising system are determined by the nature of the accessible micro-states. Thus, it is useful to know the dependence of the Hamiltonian on the spin configurations. The total energy is given by
\begin{eqnarray}
H = -J \sum^{N}_{i,j = nn(i)} s_i s_j - \mu h \sum^{N}_{i=1} s_i\,,
\end{eqnarray} 
where $J$ is the self-interaction between neighbouring spins, $h$ the external magnetic field and $\mu$ is the atomic magnetic moment. Note that in the first sum, the notation $nn(i)$ represents nearest-neighbour pairs; the sum is taken over all nearest-neighbouring pairs. The sign of $J$ determines whether we have a ferromagnetic($J = 1$) or an anti-ferromagnetic($J=-1)$ system.

In the case of the canonical ensemble, in other words, when the system is attached to a thermal reservoir and kept at a constant temperature $T$, as the time passes the spins are left to fluctuate with rates depending on the reservoir's temperature. This behaviour can be captured in a Monte Carlo (MC) simulation in the canonical ensemble. In the ferromagnetic case at $T=0$, the system is frozen with all spins being at one direction either down or up. On the other hand in the anti-ferromagnetic case at $T=0$, the system gets split into two sub-systems in a checkerboard pattern, and the difference between the spins of these two sub-systems points at one direction either up or down. The orientation of the spins is arbitrary, however, the dynamics enforce the system to choose one of the two directions. This corresponds to the spontaneous symmetry breaking of the $\mathbb{Z}_2$ global symmetry group in the ferromagnetic case. In the anti-ferromagnetic case, the existence of a checkerboard pattern corresponds to the spontaneous breaking of the translation symmetry. Although the Hamiltonian of the system is invariant under $\mathbb{Z}_2$ and translation transformations, the degenerate ground states are not invariant but get interchanged under such transformations. 

For small, nonzero values of the temperature, spins of the whole system (in the ferromagnetic case), or the sub-systems (in the anti-ferromagnetic case), still form large sectors where all spins are correlated and point to one direction. Above the critical temperature of $T_c$, the spins are disordered and $\mathbb{Z}_2$ symmetry is restored.

The question that we address in this work is whether the behaviour described above can be captured by a deep learning autoencoder when we pass it ensembles for a sequence of temperatures separated by some $\delta T$. More precisely, we seek to understand if a qualitative description of the phase structure of the Ising model can be extracted and whether one can determine the critical temperature $T_c$.

\subsection{\label{sec:swendsen-wang} Swendsen-Wang algorithm}

The MC simulation for the 2D-Ising model is conventionally performed using the Metropolis algorithm. Since this algorithm is based on local updates, it faces the problem of critical slowing down near the critical temperature, where the correlation length diverges. In order to tackle this problem, we have implemented the Swendsen-Wang cluster algorithm~\cite{Swendsen:1987ce,WANG1990565}, which is based on global updates of the spin configurations. This algorithm relies on the formation of bonds between every pair of nearest neighbours($ij$) that are aligned at a given temperature $T$, with a probability $p_{ij} = 1- \exp{ ( -2\beta J )}$, where $\beta = \frac{1}{k_BT}$ ($k_B \equiv$ Boltzmann constant). A single cluster is defined as all the spins, which are connected via bonds. The global update is defined as the collective flipping with a probability of $1/2$, on all the spins in each cluster~\cite{1969PSJJS..26...11K,Fortuin:1971dw}. This step works because of the so-called Fortuin-Kasteleyn mapping of the Ising model on the random-cluster model. Thus, global updates enable us to produce equilibrium configurations close to the $T_c$ with a few thermalization steps.

\subsection{\label{sec:monte-carlo} Monte-Carlo simulation setup}

In this work we chose to investigate the case of zero external magnetic field ($h=0$) and for simplicity we have set $J= \pm 1$ and $k_B=1$. In this case, the theoretically calculated value of the critical temperature is
\begin{eqnarray}
T_c = \frac{2}{\ln \left( 1 + \sqrt{2}\right)} = 2.269185 \,.
\label{eq:theoretically_extracted}
\end{eqnarray} 
To extract experimentally this quantity one has to investigate the order parameter of theory. The first question that we address is whether we can get an approximate estimate of this temperature  by using unsupervised learning. %
For this purpose, we define a sequence of different values of temperature. Then, for each one, we start from a "hot" configuration of spins (where the spins are oriented randomly), perform a large enough number of thermalization sweeps and then save the configuration. For every single temperature, we repeat the procedure $200$ times. {\color{black} The same results could be obtained by starting from a "cold" configuration, letting the Markov chain evolve, and then sampling configurations along the single chain, but, the former procedure guarantees a higher degree of de-correlation within the data.}\\

\noindent
\subsection{\label{sec:phase_structure} Phase structure, observables and order parameters}
The phase structure of the 2D-Ising model can be reduced to the study of the magnetic order of the system~\cite{cardy_1996,gould1988introduction}. If we suppose that there are $N_{\uparrow}$ spins pointing upwards and $N_{\downarrow}$ spins pointing downwards, then the total magnetic moment would be $N_{\uparrow} - N_{\downarrow}$ ($\mu=1$). The largest possible magnetic moment would, therefore, be $N$. Thus, for the ferromagnetic case, we can define the magnetic order parameter or magnetization per spin configuration naturally as:
\begin{eqnarray}
m=(N_{\uparrow} - N_{\downarrow})/N\, ,
\label{eq:magnetization}
\end{eqnarray}
while the average magnetization $M=\langle  m \rangle$. $M$ can get values between $-1$ and $1$, and the average of the absolute magnetization ${\tilde m} = \langle | m | \rangle$ is just the magnetic order. Hence, if ${\tilde m}$ is close to $0$, then the system is highly disordered and, thus, not magnetized, with approximately half of the spins pointing up and the other half pointing down. On the other hand, if ${\tilde m}$ is approximately $1$, the system is ordered and, thus, magnetized with nearly all the spins pointing in the same direction. 

In the anti-ferromagnetic case, the relevant magnetic order parameter is the staggered magnetization per spin configuration. In the checkerboard lattice, if we label black sites as $(+)$ and white sites as $(-)$, and we define $m_+$ and $m_-$ using (\ref{eq:magnetization}), then the staggered magnetization per spin configuration ($m_s$) can be defined as:
\begin{eqnarray}
m_s=m_{+} - m_{-}\, ,
\label{eq:stg_magnetization}
\end{eqnarray}
while the average staggered magnetization as $M_s=\langle m_s \rangle$. Similar to the ferromagnetic case, the magnetic order is the average of the absolute staggered magnetization ${\tilde m}_s = \langle | m_s | \rangle$. Therefore, if ${\tilde m_s}$ is close to $0$, then the system is highly disordered and approximately not magnetized. On the contrary, the system has exactly $0$ magnetization if ${\tilde m_s}$ is close to $1$ because in the system every spin is surrounded by the opposite spin among its neighbours, which makes it exactly ordered.

The point $T=T_{c}$ is called the {\it critical point} and separates the ordered $T<T_c$ phase and disordered $T > T_c$ phase. At $T=T_c$ the system is described by a second order phase transition, i.e. {\it \`a la} Ehrenfest~\cite{jaeger1998ehrenfest} the first derivative of the free energy with respect to the external field {which is the order parameter} is continuous while the second derivative of the free energy is discontinuous. Since there exists a bijective map between the spin fields of the ferromagnetic and anti-ferromagnetic cases of the 2D-Ising model, the phase transition in both the cases is identical.

\section{Deep Learning Autoencoders}
\label{sec:autoencoders}

\begin{figure*}[ht!]
{\rotatebox{0}{\includegraphics[width=17cm]{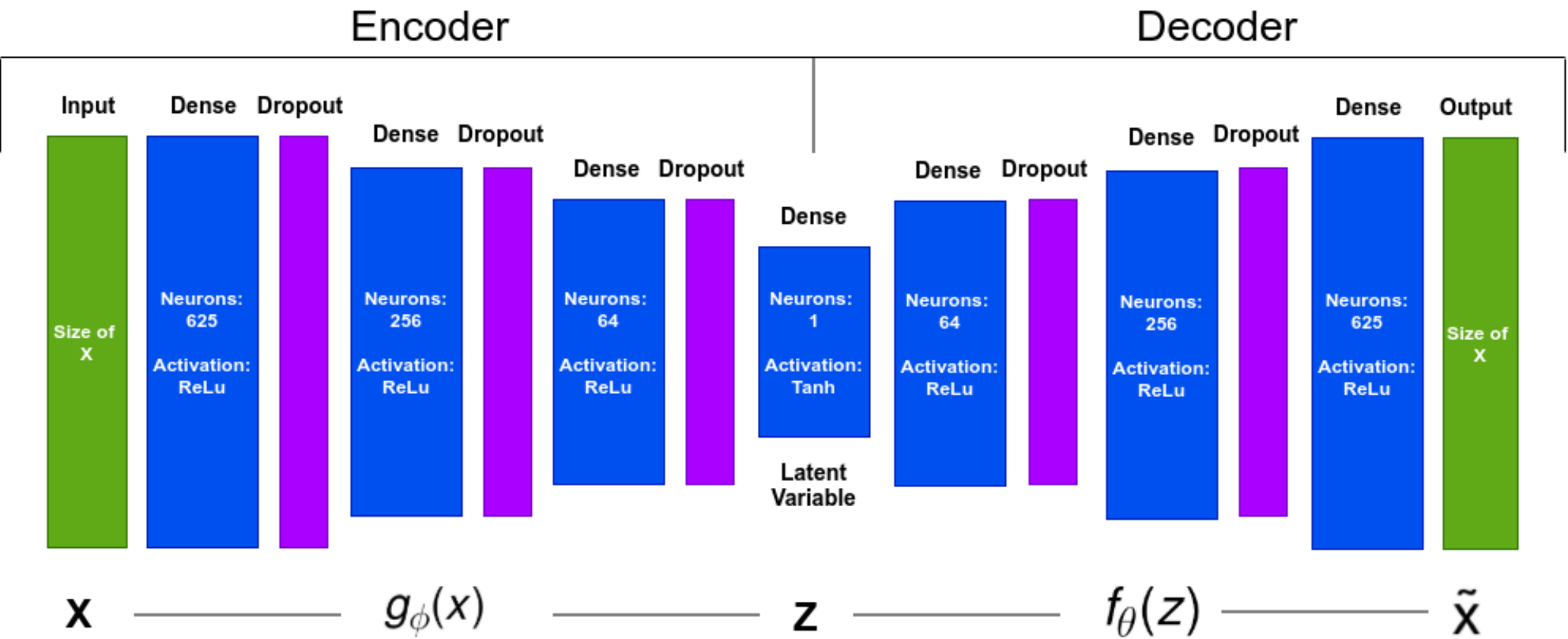}}}
\centering
\caption{\color{black} The detailed structure of the autonecoder network used in this work. The size of the input and the output layer $x$ is equal to the size of the lattice i.e. $x=L \times L$. This means that while we change the size of the lattice $L$ we keep the number of neurons and type of activation functions unaltered.}
\label{fig:autoencoder_basic_structure}
\end{figure*}

The concept of autoencoders exists for decades \cite{bourlard1988auto, hinton1994autoencoders}, where conventional autoencoders were used for feature learning and dimensionality reduction. In recent years, work has been conducted to join autoencoders and probabilistic latent variable models: Alternate forms of autoencoders have become popular for so-called generative modelling~\cite{le2008representational, kingma2013auto}. Autoencoders {\color{black} are} a variant of artificial neural networks utilized for learning data codings in an unsupervised manner, efficiently \cite{vincent2008extracting,vincent2010stacked}. An autoencoder aims to define a representation (encoding) for an assemblage of data, usually performing dimensionality reduction. An autoencoder encodes the input data ($\{ X \}$) from the input layer into a latent {\color{black} variable} ($\{ z \}$), and then uncompresses that latent {\color{black} variable} into an approximation of the original data ($ \{ \,X \}$). The autoencoder engages in dimensionality reduction by learning how to ignore the noise and recognize significant characteristics of the input data. As Fig.~\ref{fig:autoencoder_basic_structure} shows, an autoencoder consists of two components, the encoder function $g_\phi$ and a decoder function $f_\theta$ and the reconstructed input is $\,X=f_\theta(g_\phi(x))$. The first layer of an autoencoder might learn to encode simple, identifiable and local features. The second layer by using the output of the first layer learns to encode more complex and less local features. This continues for higher-order layers until the final layer of the encoder learns to identify and encode the most complex and global characteristics of the input data. The same process in reverse is true for the decoder where the goal is to go from the compressed latent variable to the original input.

{\color{black} The activation functions have been chosen in such a way so that the latent variable provides as sharp as possible transition at the assumed critical point. In other words, we investigated which combination of activation functions leads to a steeper change on the scattered latent variable (like for instance figure~\ref{fig:autoencoder_model}), in order to identify with as much accuracy as possible the critical temperature for a given lattice volume. Other combinations of activation functions have also been investigated and will be presented in a forthcoming work ~\cite{Alexandrou:inpreparation}}.

In the training phase, the autoencoder learns the parameters $\phi$ and $\theta$  together, where $f_\theta(g_\phi(x))$  can approximate an identity function. Various metrics can be used to measure the error between the original input $X$ and the reconstruction ${\tilde X}$, but the most simple and most commonly used is the Mean Square Error (MSE) as this is provided in Eq.~\ref{eq:MSE}, where $n_{\rm data}$ is the number of data points:
\begin{equation}
MSE(\theta, \phi) = \frac{1}{n_{\rm data}}\sum^{n_{\rm data}}_{i=1}(X_i - f_\theta(g_\phi(X_i)))^2\,.
\label{eq:MSE}
\end{equation}

\subsection{Proposed Autoencoder Model}
\label{sec:proposed_autoencoder}

For the analysis of the proposed method, an eight-layer, fully connected (Dense), autoencoder is proposed, as Fig.~\ref{fig:autoencoder_basic_structure} shows, where the encoder compresses the configurations into a single latent {\color{black} variable}. Through experimentation, we determine that the best model to detect the transition consists of the encoder with the input layer, first, second and third hidden layers having 625, 256, 64 and 1 neurons, respectively. {\color{black} The input layer has size equal to $L \times L$ and, thus, it changes as we alter the lattice size}. The activation function used is ReLu (rectified linear unit), as shown in Eq.~\ref{eq:activation_relu}, for all layers except the third hidden layer, where tanh was used, as shown in Eq.~\ref{eq:activation_tanh}. For the decoder, the first, second and third hidden layers use 64, 256, and 625 neurons, respectively. For the output layers, the number of neurons is set to be equal to the number of lattice points in the configuration under investigation. The activation function used is ReLu, as given in Eq.~\ref{eq:activation_relu}, for all hidden layers, and for the output layer, tanh is used, as per Eq.~\ref{eq:activation_tanh}.
\begin{equation}
{\rm ReLu}: y = max(0, x) = \left\{ \begin{array}{ll}  x, & \mathrm{if} \ x  > 0 \\  0 & 
\mathrm{if}  \ x \leq 0   \end{array} \right\}. \ 
\label{eq:activation_relu}
\end{equation}

\begin{equation}
{\rm tanh}: y = \frac{1-e^{-2x}}{1+e^{-2x}} \,.
\label{eq:activation_tanh}
\end{equation}

For the proposed autoencoder model, we use the so-called dropout realization technique~\cite{hinton2012improving}, on 30\% of the neurons at each layer. The dropout regularization technique refers to temporarily deactivating neurons from each layer, randomly, when training. It was successful at reducing over-fitting in our case. For the training of the proposed autoencoder model, the data are split into training (66.66\dots\%) and testing (33.33\dots\%) sets and the training is performed for 2000 iterations. For training, the starting learning rate was set to 0.001 and was reduced by 20\%, when learning stagnates for 30 epochs, with minimum learning rate 0.000001. The implementation was performed using Keras \cite{chollet2015keras} and Tensorflow \cite{abadi2016tensorflow}.

\section{Results for the ferromagnetic Ising model}
\label{sec:results}

\subsection{The latent {\color{black} variable} per configuration}
\label{subsec:latentforconfiguration}

In order to identify signals of the phase structure of the 2D-Ising ferromagnetic model, as a first step, we investigate how the latent {\color{black} variable} $z_{i_{\rm conf}}$ behaves as a function of the temperature $T$ for each configuration. We produce $40000$ configurations, namely $200$ configurations for every single temperature. The produced configurations are for $200$ different values of temperatures within the range $T=1-4.5$ and separated by $\delta T = 0.0175$. We choose this range to make sure that we cover the two extreme cases, the nearly "frozen" at $T \simeq 1$ and completely disordered $T \simeq 4.5$. Furthermore, we assume that we have no prior knowledge of what is happening in between these two extremes. 

After training the proposed autoencoder on configurations of the 2D-Ising model, the reconstruction error was found to be relatively high ($0.6-0.7$) with both training and testing sets. {\color{black} We clarify that the autoencoder is trained for all temperatures together in one dataset for each lattice size $L$.} Table \ref{MSE_errors} shows the results. As the configurations consist of spins with values of 1 and -1, the maximum possible error is 4 based on MSE.

\begin{table}[h]
\centering
\normalsize
\begin{tabular}{cccc} \hline
$L$              & \textbf{Training Error} & \textbf{Testing Error} & \textbf{ARA} \\ \hline
{25}  & 0.6136           & 0.6189         & 37.30\% \\
{35}  & 0.6346           & 0.6408         & 36.70\%\\
{50}  & 0.6637           & 0.6688         & 35.50\%\\
{100} & 0.6764           & 0.6822         & 33.70\%\\
{150} & 0.6946           & 0.6994         & 31.90\% \\ \hline   
\end{tabular}
\caption{MSE Traning and Testing Errors of the Reconstructions Based on the Proposed Model as well as the Average Reconstructed Accuracy (ARA).}
\label{MSE_errors}
\end{table}

\begin{figure}
{\includegraphics[width=6.8cm]{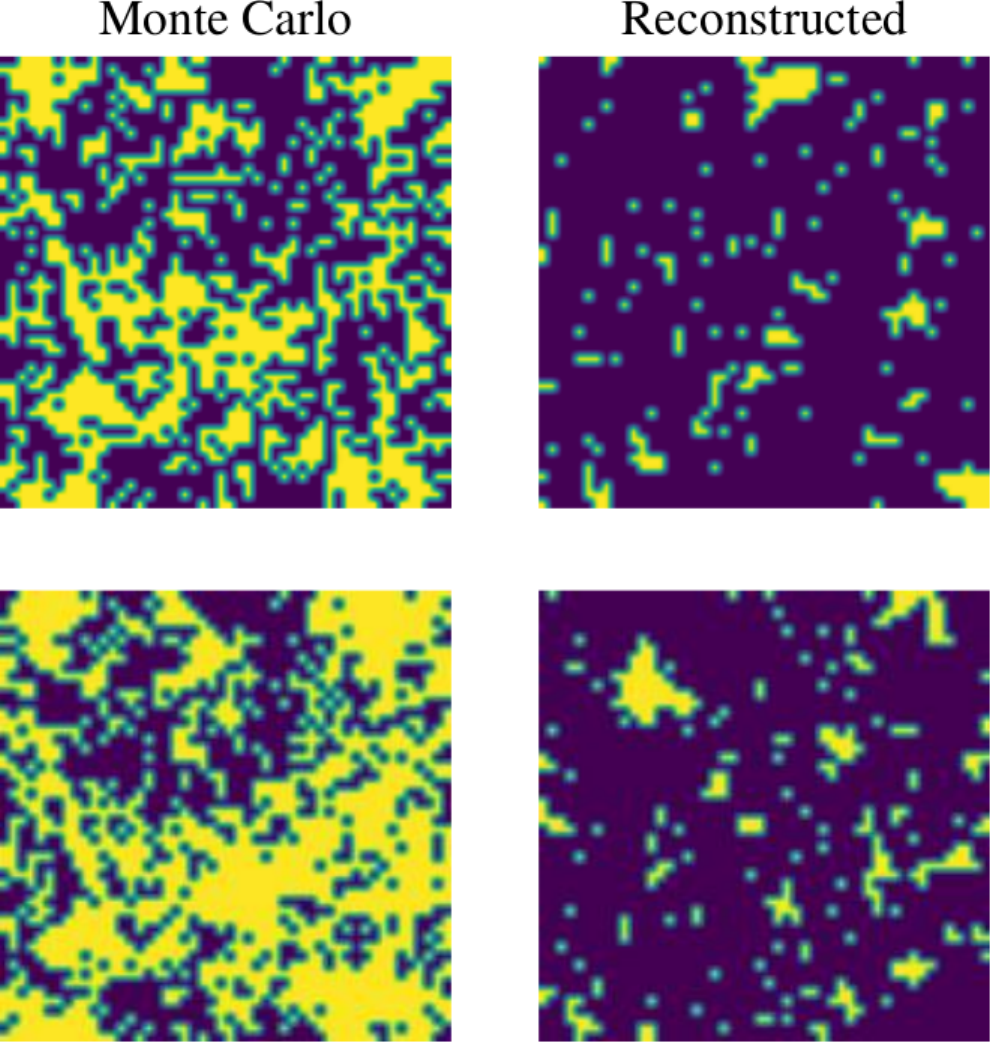}\put(-215,145){\large $T_1$}\put(-215,45){\large $T_2$}}
\centering
\caption{\color{black} Examples of reconstructed configurations and their comparison to the actual Monte-Carlo calculations from which the latent variable originated. The first line corresponds to $T_1=2.2$ and the second line to $T_2=2.27$. Blue area corresponds to spin "up" regions and yellow area to spin "down".}
\label{fig:recontruction}
\end{figure}

\begin{figure*}[ht!]
  \begin{center}
\includegraphics[width=13cm]{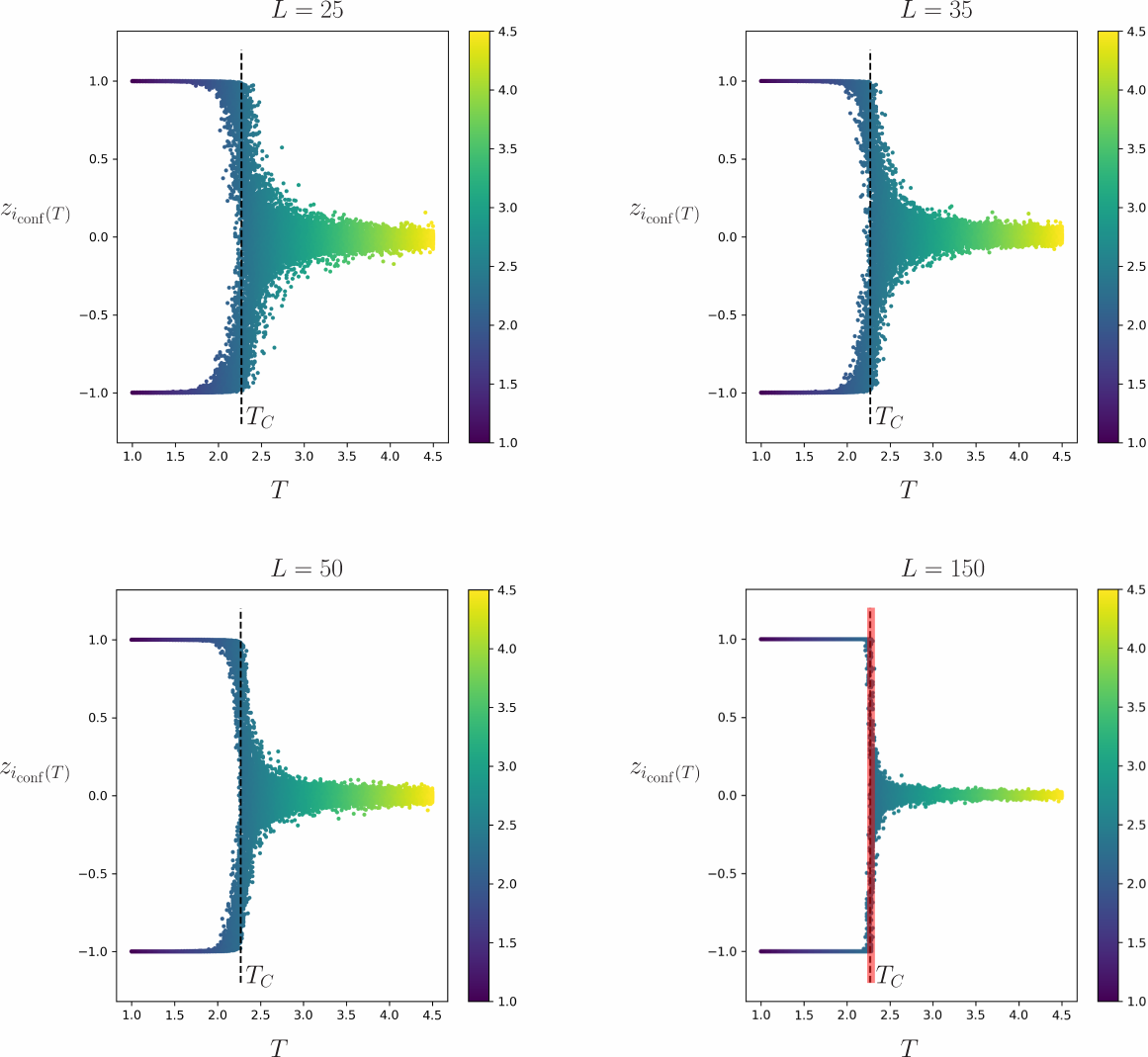}
  \end{center}
\caption{The latent {\color{black} variable} for each configuration as a function of the temperature for four different lattice volumes. {\color{black} These are scatter plots where no averaging was done for every single input data}. The dashed line represents the analytically extracted value of the critical temperature (Eq.~\ref{eq:theoretically_extracted}). The red shaded area in the plot for $L=150$ is the region where (by fitting to a constant) we expect to find the $T_c(L=150)$. The color on the gradient illustrator on the right denotes the temperature $T$.}
\label{fig:autoencoder_model}
\end{figure*}

It appears that both training and testing errors increase with the lattice size. This was expected as the dimensionality (the number of configuration components) increases. {\color{black} Table~\ref{MSE_errors}, also provides the average reconstructed accuracy, which appears to be low and to decrease with the size of the lattice. This indicates the complexity of the problem in terms of reconstruction from the encoded variables as the error increases with the size of the configurations. A crucial point is the very low number of latent variables, in this case, only one. This is also demonstrated in Fig.~\ref{fig:recontruction}, where we present the result of reconstruction for two configurations at two different temperatures close to the critical point and for lattice size $L=50$. This result resembles the poor reconstruction accuracy observed in Ref.~\cite{Cristoforetti:2017naf} when using variational autoencoder to reconstruct configurations for the 2D XY model. The reconstruction error appears to be small for low temperatures with an average MSE error of $\sim 0.015$ at $T=1$. As the temperature increases, the error rises until it reaches $T_c$ for which it becomes $\sim 1$ and stops changing further with the temperature.}
In Fig.~\ref{fig:autoencoder_model} we show the latent {\color{black} variable} for each different configuration, as a function of the temperature $T$, for four different lattice sizes, $L=25$, $35$, $50$, $150$. 

\begin{figure}[!ht]
  \begin{center}
  \vspace{-0.5cm}
    {{\includegraphics[width=8cm]{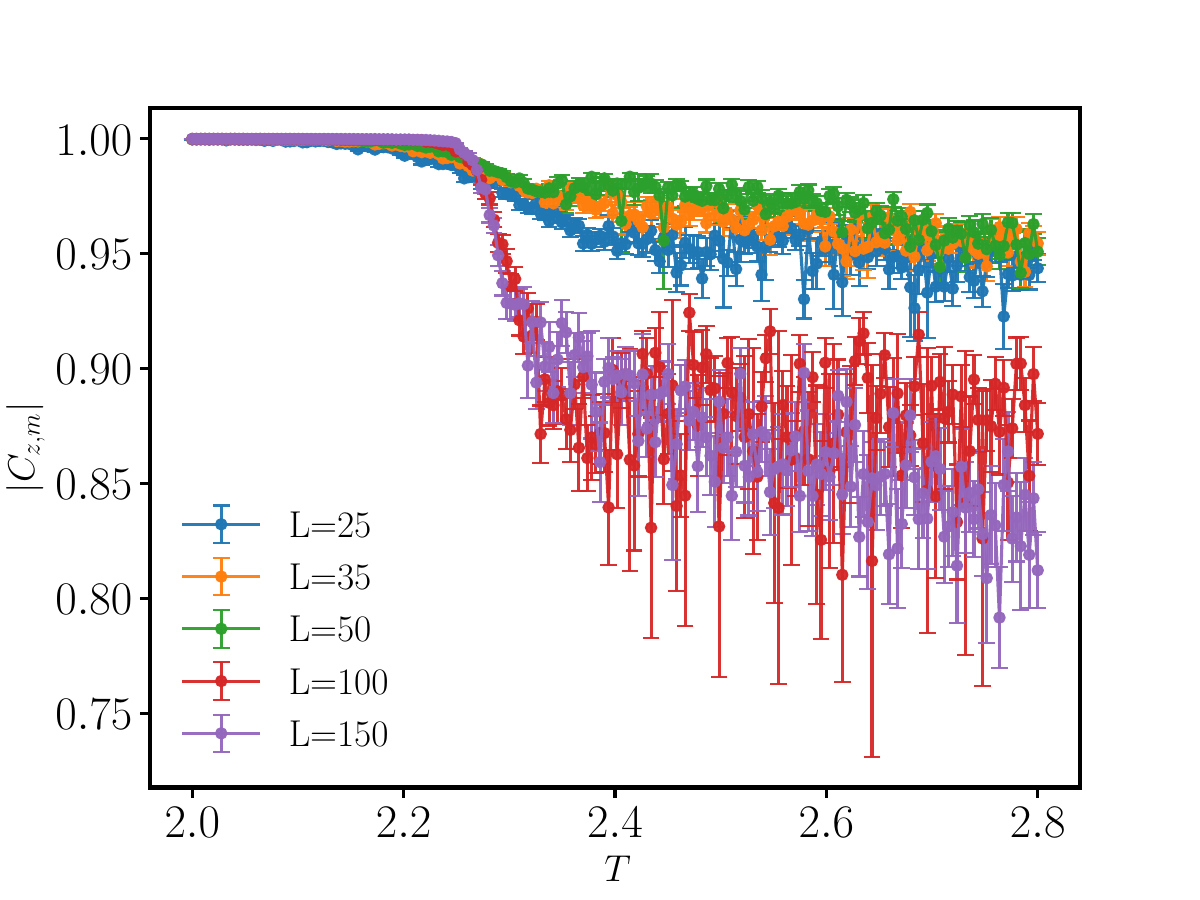}}}
  \end{center}
\caption{The absolute correlation coefficient defined in Eq.~\ref{correlation}  as a function of the temperature for the five different lattice sizes.}
\vspace{-0.5cm}
\label{fig:correlation_coefficient}
\end{figure}

Fig.~\ref{fig:autoencoder_model} has the following features:

\begin{itemize}
    \item For low temperatures, we obtain two plateaus, one located at $z=1$ and one at $z=-1$. A first simplistic explanation for this pattern would be that it corresponds to two distinct states that are not connected through any kind of transformation. This reflects the spontaneously broken $\mathbb{Z}_2 \equiv \{-1,1\}$ global symmetry group. One can interpret these two plateaus as the two cases where all spins are up or down. This interpretation is confirmed by the results presented in Fig.~\ref{fig:correlation_coefficient} where we show the absolute correlation coefficient $C_{z,m}$ between the latent {\color{black} variable} $z$ and the magnetization $m$ defined as 

    \begin{eqnarray}
      C_{z,m} = \frac{\langle (z - {\bar z})(m - {\bar m}) \rangle}{\sqrt{\langle (z - {\bar z})^2 \rangle \langle (m - {\bar m})^2 \rangle}}\,.
      \label{correlation}
    \end{eqnarray}
    The fact that at low temperatures the absolute correlation coefficient is $1$ demonstrates that the two different values of the latent {\color{black} variable} $-1$ and $1$ correspond to the two orientations of the spins. Finally, the two plateaus become more distinct as the lattice size increases.

    \item At some temperature range $\Delta T_{\rm trans}$ the aforementioned behaviour collapses to one state, which is located around $z=0$. This reflects the restoration of $\mathbb{Z}_2$ symmetry. In other words, it corresponds to the case where all the spins are disoriented. 

    \item There is a critical point where there is a change in the pattern. As the lattice size increases the width of this transition decreases with $\Delta T_{\rm trans} \to 0$ and this step becomes steeper and steeper. At $L \to \infty$ the transition is localized right on the critical temperature $T_c$ extracted analytically. 
\end{itemize}
Evidently, plotting the latent {\color{black} variable} as a function of the temperature demonstrates that the autoencoder "notices" the two phases. Also, it provides a good approximation of the critical temperature. In fact from Fig.~\ref{fig:autoencoder_model} the transition appears to occur right at the critical temperature for $L=150$. We fit the points of the latent {\color{black} variable} which, to a good approximation, behave linearly to a constant as a function of $T$. This enables us to restrict that the collapse of the two states located at $1$ and $-1$ occurs at $T \simeq 2.28(4)$. This {\color{black} temperature} region is denoted in Fig.~\ref{fig:autoencoder_model} as the shaded area in red. In Fig.~\ref{fig:correlation_coefficient} we observe that within this {\color{black} temperature} region the value of the absolute correlation coefficient $C_{z,m}$ starts to decrease from $1$. This demonstrates that, although highly correlated, the latent {\color{black} variable} and magnetization are two different quantities. {\color{black} This result shares similarities with the findings of Ref.~\cite{wetzel2017unsupervised} where the authors showed that the latent variable resulting from an autoencoder and a variational autoencoder has some level of correlation with the order parameter. The authors used an autoencoder with one fully connected hidden layer with 256 neurons and ReLu activation function and a final layer with a sigmoid activation function.} {\color{black} A simple explanation of the ability of the autoencoder to detect the phase transition is due to the fact that the variances of data which is fed as input are becoming maximal at the point of the phase transition. A possible explanation of the behaviour of the latent variable of being steeper than that of the magnetization close to the critical point is that the choice of non-linear activation function in the network "distorts" and possibly enhances certain contributions to the variances. Namely, we have observed that different choices of activation functions would affect the steepness of the average latent variable. As a matter of fact, this combination of activation functions maximizes the steepness of the average latent variable close to the critical point. A comprehensive study of how different choices of layers, number of neurons and activation functions are affecting the behaviour of the latent variable is under way~\cite{Alexandrou:inpreparation}}.

\begin{figure}[H]
  \begin{center}
  \vspace{-1.5cm}
\rotatebox{0}{\includegraphics[width=9cm]{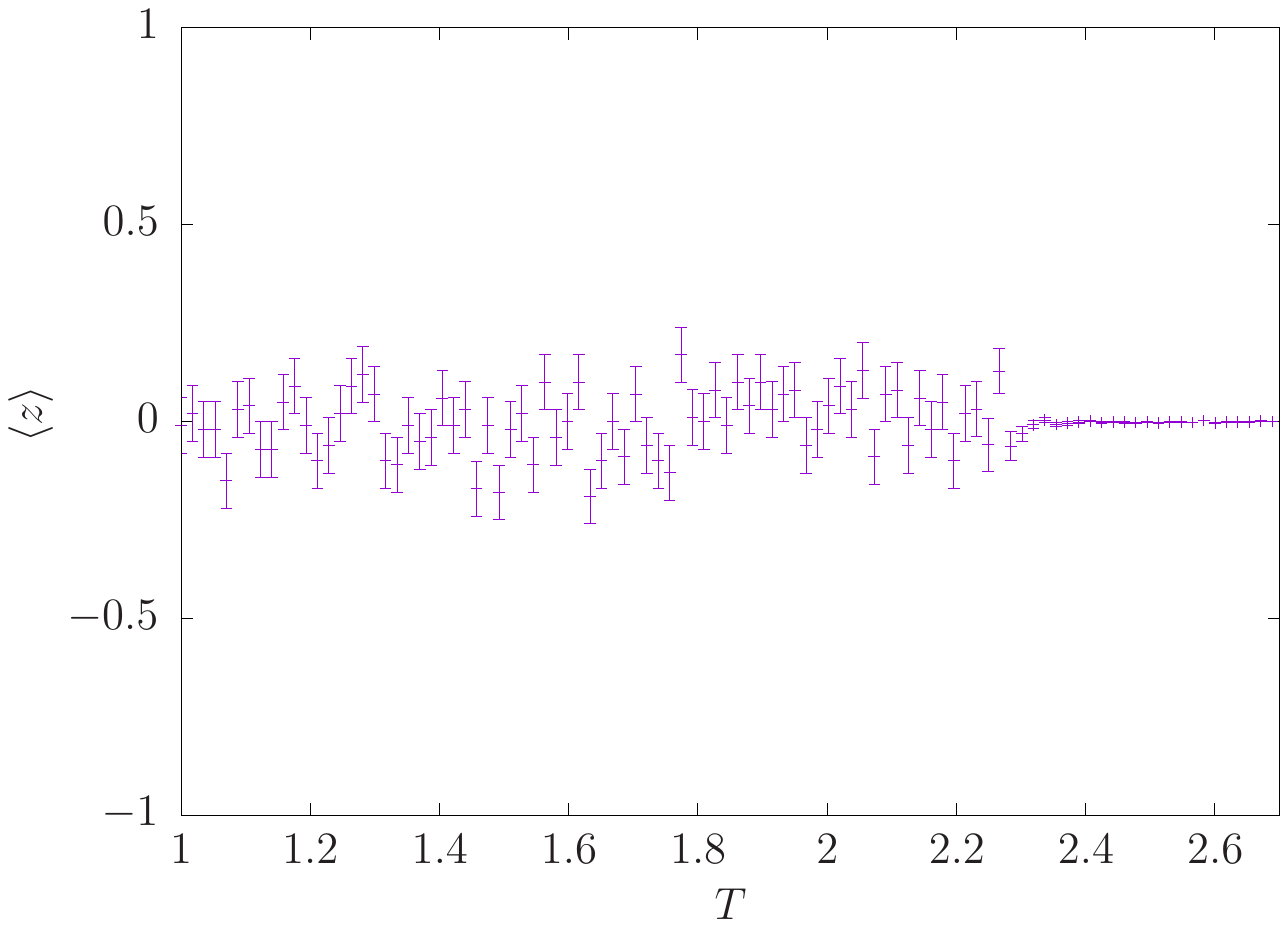}}
\vspace{-1.0cm}
  \end{center}
\caption{\color{black} The average latent {\color{black} variable}  $\langle z \rangle$ as a function of the temperature $T$ for $L=150$. This demonstrates that the latent {\color{black} variable} is equally distributed around zero for all values of the critical temperature $T_c$. }
\label{fig:distribution}
\end{figure}

Finally, we observed that for low temperatures the latent {\color{black} variable} is, in a good approximation, equally distributed between the values of $z=1$ and $z=-1$. This can be seen in Fig.~\ref{fig:distribution} where we present the average latent {\color{black} variable} $\langle z \rangle$ for each temperature and $L=150$ as a function of the temperature. 
\begin{figure*}[!ht]
\hspace{1.5cm} 
\includegraphics[width=7.5cm]{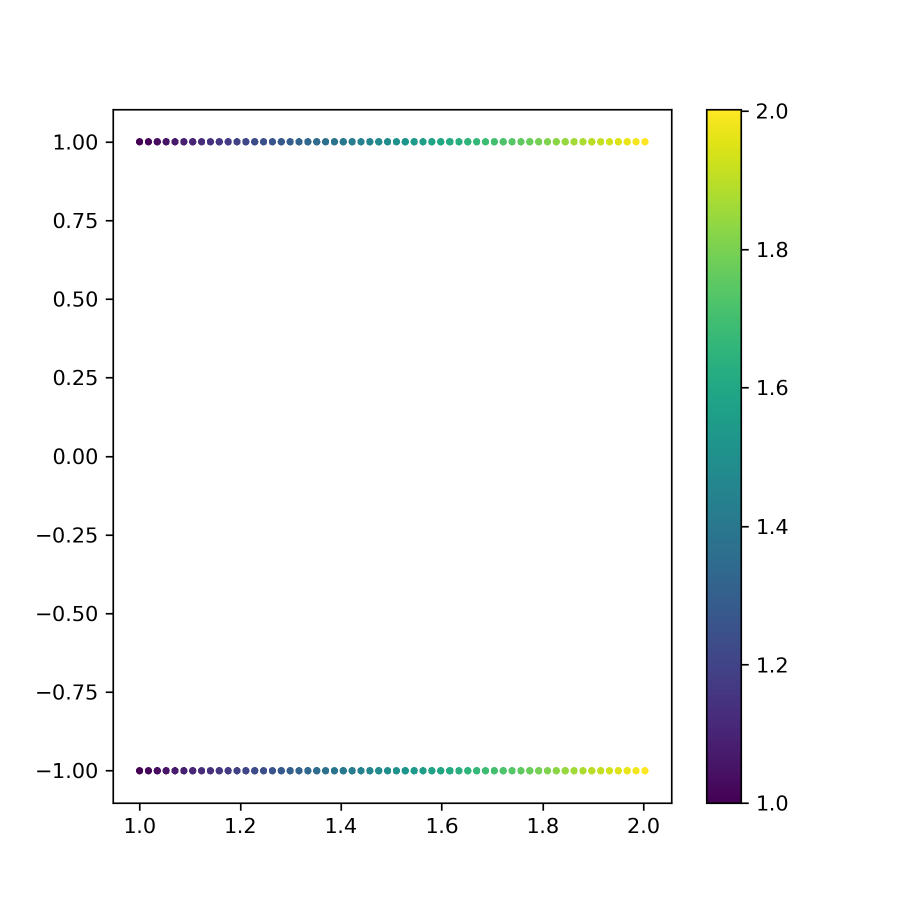} \put(-130,0){\large $T$}
\put(-235,120){\large $z_{i_{\rm conf}}$}
\hspace{0.5cm}
\includegraphics[width=7.5cm]{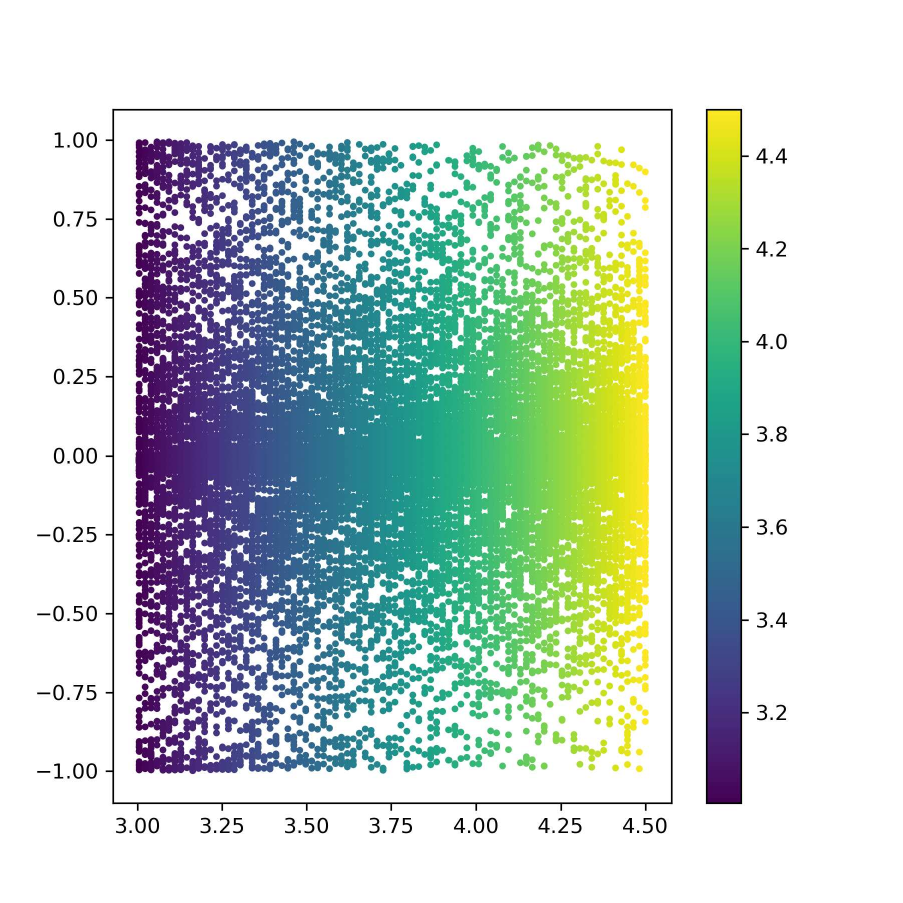} 
\put(-130,0){\large $T$}
  \hspace{-4cm}
\caption{The latent {\color{black} variable} for each configuration as a function of the temperature for $L=100$, obtained by applying the autoencoder on configurations produced within the range of temperatures $T=1-2$ (left) and $T=3-4.5$ (right). {\color{black} Mind that the critical temperature $T_c=2.26918$ is out of range in the two plots above since no data produce around this temperature have been included as input in the autoencoder.} }
\label{fig:autoencoder_model_2}
\end{figure*}

One could also investigate what happens within different "temperature windows". For instance, we can use a temperature window within the range $T=1-2$ and apply the autoencoder. The outcome would be the behaviour presented in the left panel of Fig.~\ref{fig:autoencoder_model_2}, where only the two ordered states are visible without the presence of a critical point. Since there is no visible signal for a phase transition behaviour within this range of $T$, it is reasonable to use another temperature window. If we choose  $T=3-4.5$, for instance, the corresponding latent {\color{black} variable} would be the one given on the right panel of Fig.~\ref{fig:autoencoder_model_2} where no particular pattern is observed. A sensible next step would be to investigate what happens within a range of temperatures located between the two previous temperature windows, for instance, $T=1-4$.

\subsection{The absolute average latent {\color{black} variable}}
\label{subsec:absolutelatentdimension}

Since the latent {\color{black} variable} per configuration is symmetric with respect to the $T$ axis, it would be reasonable to define the average absolute latent {\color{black} variable} as a parameter indicating the phase as 
\begin{eqnarray}
{\tilde z}=\frac{1}{N_{\rm conf}} \sum_{i=1}^{N_{\rm conf}} \left| z_{i_{\rm conf}} \right| \,.
\label{eq:average_absolute_latent}
\end{eqnarray}
 Fig.~\ref{fig:autoencoder_model} shows that the latent {\color{black} variable} resembles the behaviour of the magnetization per spin configuration as a function of the temperature. The absolute average magnetization defines the order parameter of the system distinguishing the two different phases. For the case of the autoencoder we can define an additional quasi-order parameter as the absolute average latent {\color{black} variable}.

\begin{figure*}[!ht]
  \begin{center}
\rotatebox{90}{\includegraphics[width=7cm]{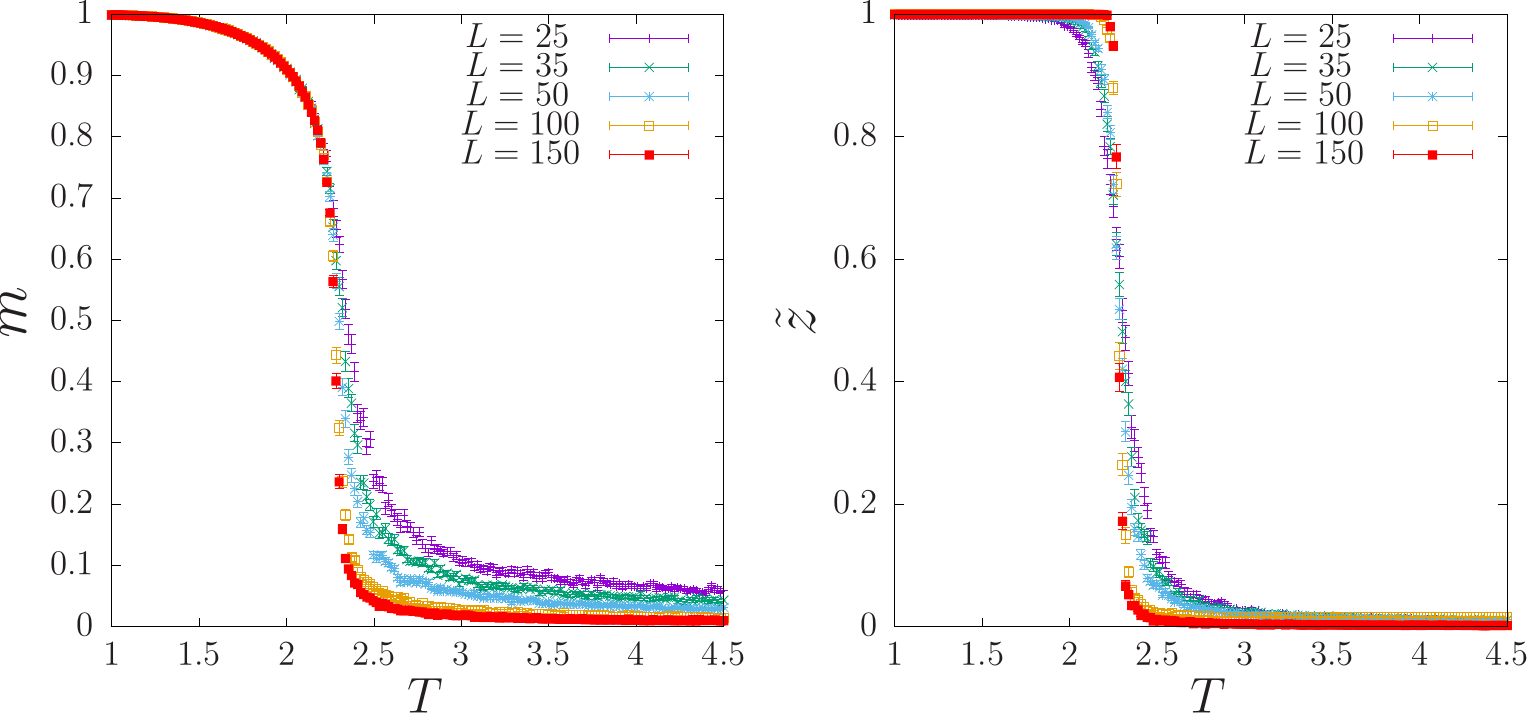}}
  \end{center}
\caption{\color{black} The average absolute magnetization (left) compared to the average latent {\color{black} variable} (right) as a function of the temperature for five different lattice volumes. Magnetization has a behaviour in accordance with a second order phase transition while the average latent {\color{black} variable} appears to have behaviour resembling a first order phase transition. The data presented on the right plot stem from testing data. {\color{black}Each data point has been extracted as an ensemble average of the observable for fixed volume and temperature.}}

\label{fig:average_latent}
\end{figure*}

In the left-hand side of Fig.~\ref{fig:average_latent} we provide the magnetization as a function of the temperature while on the right-hand side we provide the absolute latent {\color{black} variable}. Indeed the absolute latent {\color{black} variable} looks similar to the magnetization, albeit becoming steeper as the lattice size increases. Clearly, the magnetization behaves as an order parameter with the characteristics of a second order phase transition while the absolute latent {\color{black} variable} resembles a behaviour consistent with a first-order phase transition. Of course, a more careful study needs to be carried out in order to understand better whether the latent variable can actually capture a first-order phase transition. We can, therefore, conclude that the absolute average latent {\color{black} variable} can be used as an order parameter to identify the critical temperature, but cannot capture the right order of the phase transition. The fact that ${\tilde z}$ as a function of the temperature becomes steeper as the lattice size increases suggest that the critical temperature $T_c(L)$ as a function of the lattice size $L$ extracted from the autoencoder data will suffer less from finite-size scaling effects as discussed in detail in Section \ref{subsec:criticaltemperature}.

Traditionally,  $T_c(L)$ can be extracted by probing the peak of the magnetic susceptibility $\chi$ at zero magnetic field $h$, where
\begin{eqnarray}
\chi = \frac{L^2}{T} \left( \langle m^2 \rangle - \langle m \rangle^2 \right)\,.
\end{eqnarray}
According to finite size scaling theory, close enough to $T_c$, magnetic susceptibility $\chi$ scales as
\begin{eqnarray}
\chi \propto \left( t \right)^{-\gamma}\,,
\end{eqnarray}
where $t=(T-T_c)/T_c$ is the reduced temperature and $\gamma=7/4$ a critical exponent~\cite{cardy_1996}. The magnetic susceptibility measures the ability of a spin to respond due to a change in the external magnetic field. In the same manner we define the {\it latent susceptibility} as
\begin{eqnarray}
\chi_{\tilde z} = \frac{L^2}{T} \left( \langle {\tilde z}^2 \rangle - \langle {\tilde z} \rangle^2 \right)\,.
\end{eqnarray}
{\color{black}Another conventional route to obtain the $T_c$ is by computing the fourth-order cumulant of the order parameter, also known as the \textit{Binder cumulant} \cite{Binder:1981zz}, defined as }
\begin{eqnarray}
U_4 = 1 - \frac{\langle m^4 \rangle}{3\langle m^2 \rangle^2}\,.
\end{eqnarray}                                        
{\color{black}This quantity aims at capturing the non-trivial fluctuations of higher order in the spin, thus, excluding the trivial Gaussian fluctuations.  In the thermodynamic limit ($L\to \infty$), the cumulant becomes $\frac{2}{3}$ for $T<T_c$ and $0$ for $T>T_c$.}
\begin{figure}[!ht]
  \begin{center}
     {\includegraphics[width=9cm]{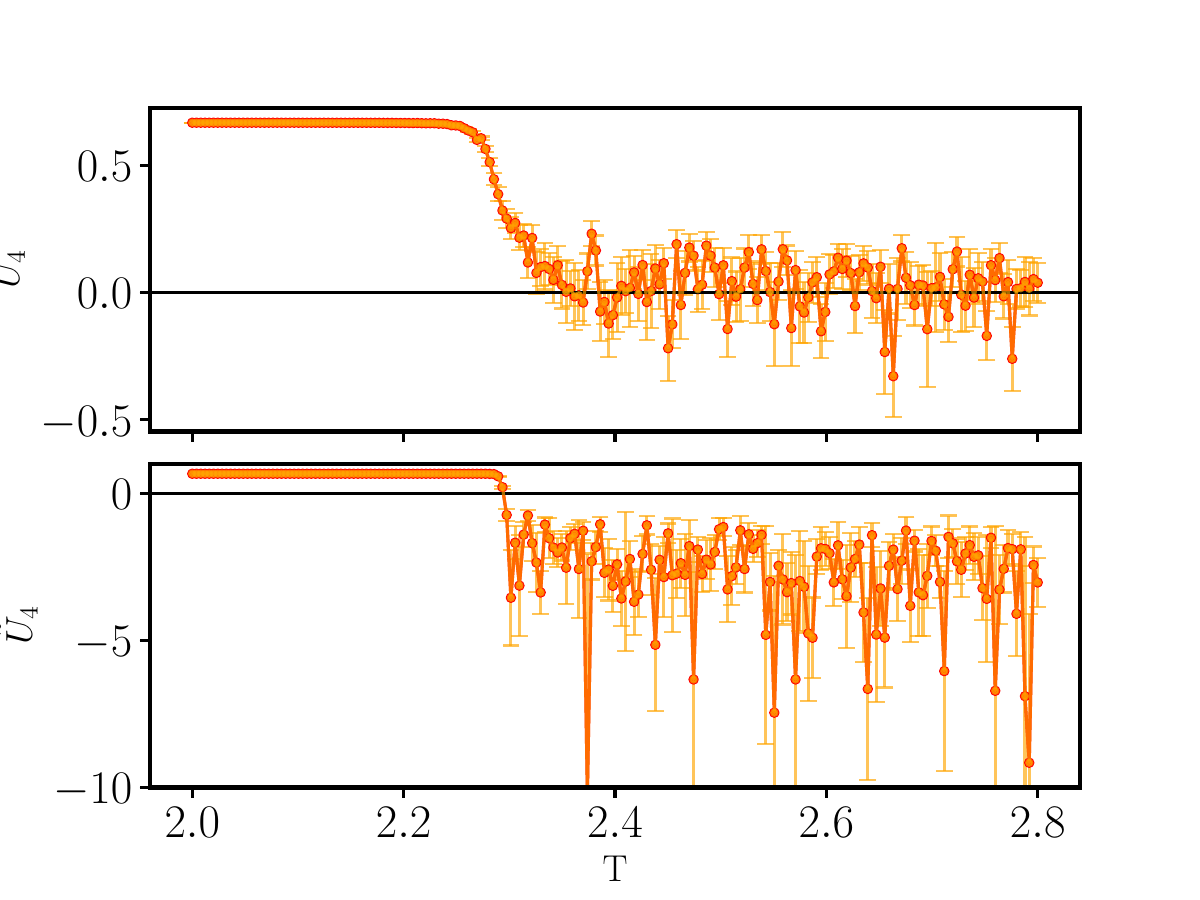}}
  \end{center}
\caption{{\color{black} The Binder cumulants $U_4$ and $\tilde U_4$ plotted as a function of temperature for $L=150$. The dotted line indicates the theoretical $T_c$ for infinite volume. }}
\label{fig:binder_cumulant}
\end{figure}

\begin{figure}[!ht]
  \begin{center}
     {\includegraphics[width=9cm]{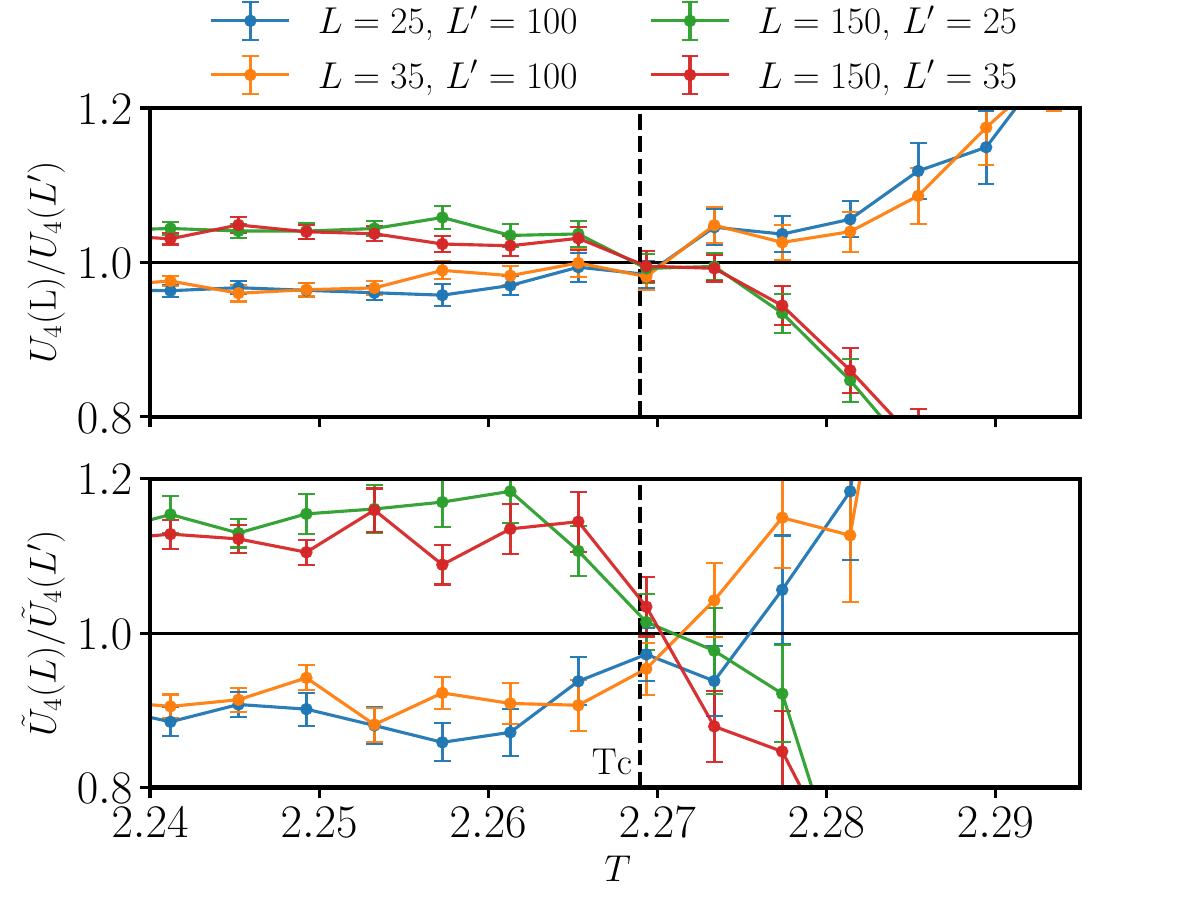}}
  \end{center}
\caption{{\color{black} The Binder cumulant ratios $\frac{U_4(L)}{U_4(L')}$ and $\frac{\tilde U_4(L)}{\tilde U_4(L')}$ are plotted with $T$.    }}
\label{fig:binder_cumulant_ratio}
\end{figure}

{\color{black} We define a similar quantity with respect to the latent variable $\tilde z$, as}     
\begin{eqnarray}
\tilde U_{4} = 1-\frac{\langle {\tilde z}^4 \rangle}{3\langle {\tilde z}^2 \rangle^2}\,.
\end{eqnarray}
{\color{black}Binder cumulants $U_4$ and $\tilde U_4$ have been plotted for the largest volume in Fig.~\ref{fig:binder_cumulant}. The $U_4$ obtained from Monte Carlo simulations is consistent with the thermodynamic limit. On the other hand, the $\tilde U_4$ obtained using the latent variable exhibits a plateau below $0$ which makes it ambiguous to comment on the order of the phase transition using autoencoders. 

Binder cumulants can be further utilised in the \textit{cumulant ratio intersection method} to extract the $T_c$~\cite{balianillcondensed}, independently of the critical exponents.  In Fig~\ref{fig:binder_cumulant_ratio}, we have demonstrated the application of this method to pinpoint the $T_c$ using $U_4$ and $\tilde U_4$. The weak dependence of the Binder cumulant $U_4$ on $L$, keeps it close to the (universal but nontrivial) \textit{fixed-point} $U_4^*$.\footnote{$U_4^* = \lim_{L\to\infty} U_4(T=T_c, L)$} Therefore at $T=T_c$, the cumulant ratios for all finite volumes intersect the $y=1$ line. The $\tilde U_4$ Binder cumulant ratio exhibits a more noisy behaviour while crossing $y=1$, which is the prime reason we resorted to the latent susceptibility method discussed in the next section to extract the $T_c$.  } 

For the extraction of $T_c$  one realizes, by looking at Fig.~\ref{fig:average_latent}, that more data points close to the critical behaviour are needed to extract the critical temperature from the latent susceptibility. Hence, we produce configurations for a grid of temperatures near the critical regime. More specifically, we produce 200 configurations per temperature, for 200 different values of $T$ in the range of $T=2-2.8$ and $\delta T=0.004$ for all the volumes considered. In addition, for $L=100$ and $L=150$ we produce 200 configuration for each  value of $T$, in the range of $T=2.22-2.34$ with $\delta T = 0.0006$. These new configurations, however, are not used to train the autoencoder. Instead, we use the synaptic weights extracted  and predict the latent {\color{black} variable} for the new configurations. Hence, this serves as a confirmation that our data do not suffer from over-fitting.
\begin{figure*}[!ht]
  \begin{center}
    \rotatebox{90}{\includegraphics[width=6cm]{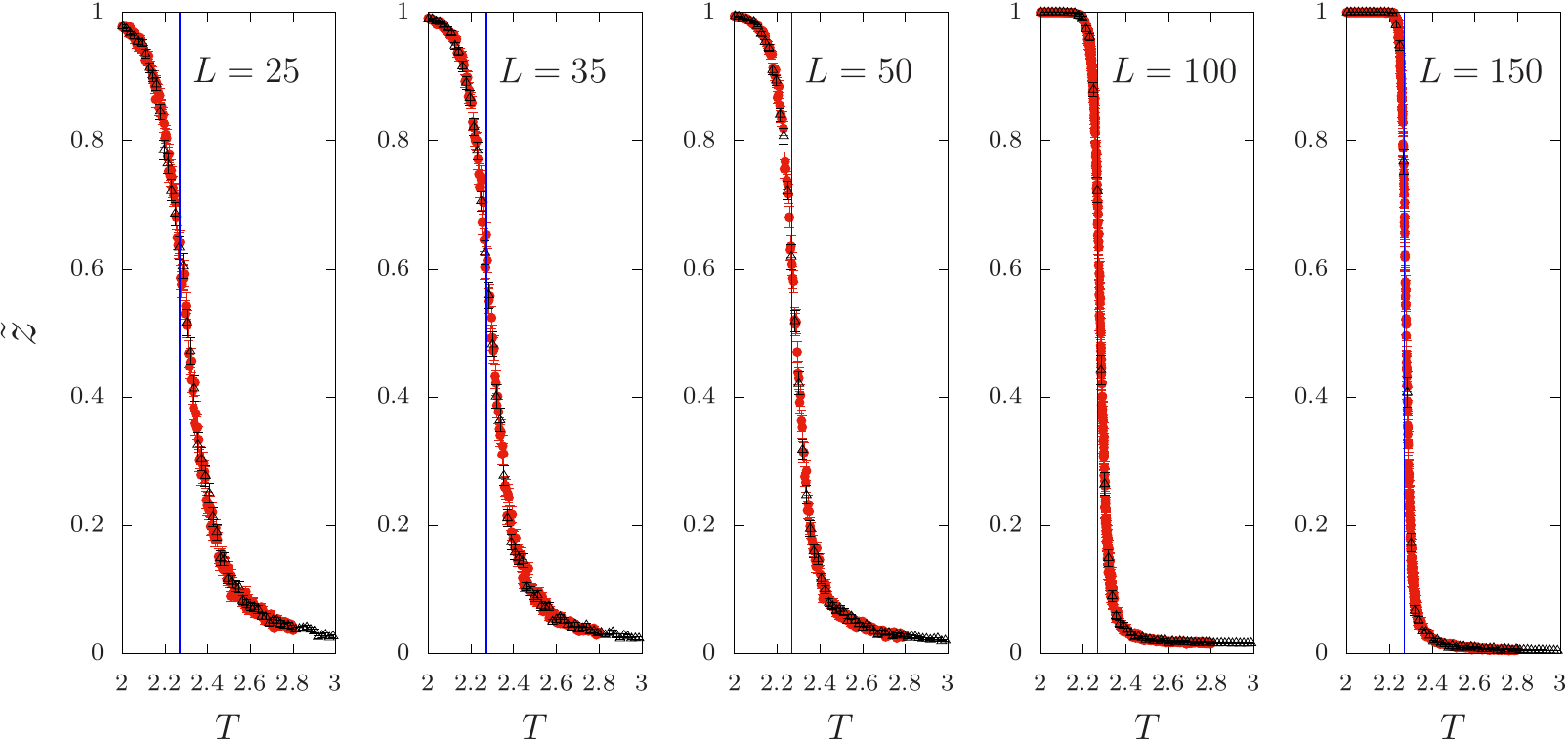}}
  \end{center}
\caption{The red points show the predicted latent {\color{black} variable} for the configurations produced around the critical point $T_c$ while in black the latent variable extracted from the first set of temperatures on which we have trained the encoder, nevertheless on test data sets. The blue vertical line corresponds to the analytically extracted critical temperature $T_c$. The figure shows that the synaptic weights extracted by training the autoencoder on the initial set of temperatures can successfully predict the latent variables for the new set of temperatures demonstrating there is no over-fitting occurring.}
\label{fig:average_latent_detailed}
\end{figure*}

In Fig.~\ref{fig:average_latent_detailed}, we present the results of applying the autoencoder weights on the new configurations produced in the region close to the critical point. We compare with results extracted using configurations produced in the range $T=1-4.5$. Both datasets agree, and there is a nice continuation of the behaviour of the absolute average latent {\color{black} variable} within the critical regime. This serves as a confirmation that the execution of the encoder does not suffer from any over-fitting occurrence and at the same time, more data points can be used for the extraction of  $\chi_{\tilde z}$. {\color{black} To avoid any confusion, we state that all generated data which are presented in figures have been extracted exclusively from the test datasets}. Furthermore, the plot for $L=150$ behaves nearly as a step function with the step being right on the theoretically extracted $T_c$. By fitting the second moment of the latent {\color{black} variable}, as this is described in section~\ref{subsec:criticaltemperature}, one sees that the transition occurs at $T_c(L=150) = 2.2779(3)$; this value is very close to the theoretically extracted value $T_c = 2.26918$.

In the following section, we present the analysis of our data in order to investigate the latent susceptibility $\chi_{\tilde z}$ and, to subsequently, extract the critical temperature $T_c(L)$ from the corresponding peak.

\begin{figure*}[!ht]
\begin{center}
    \rotatebox{90}{\includegraphics[width=7.00cm]{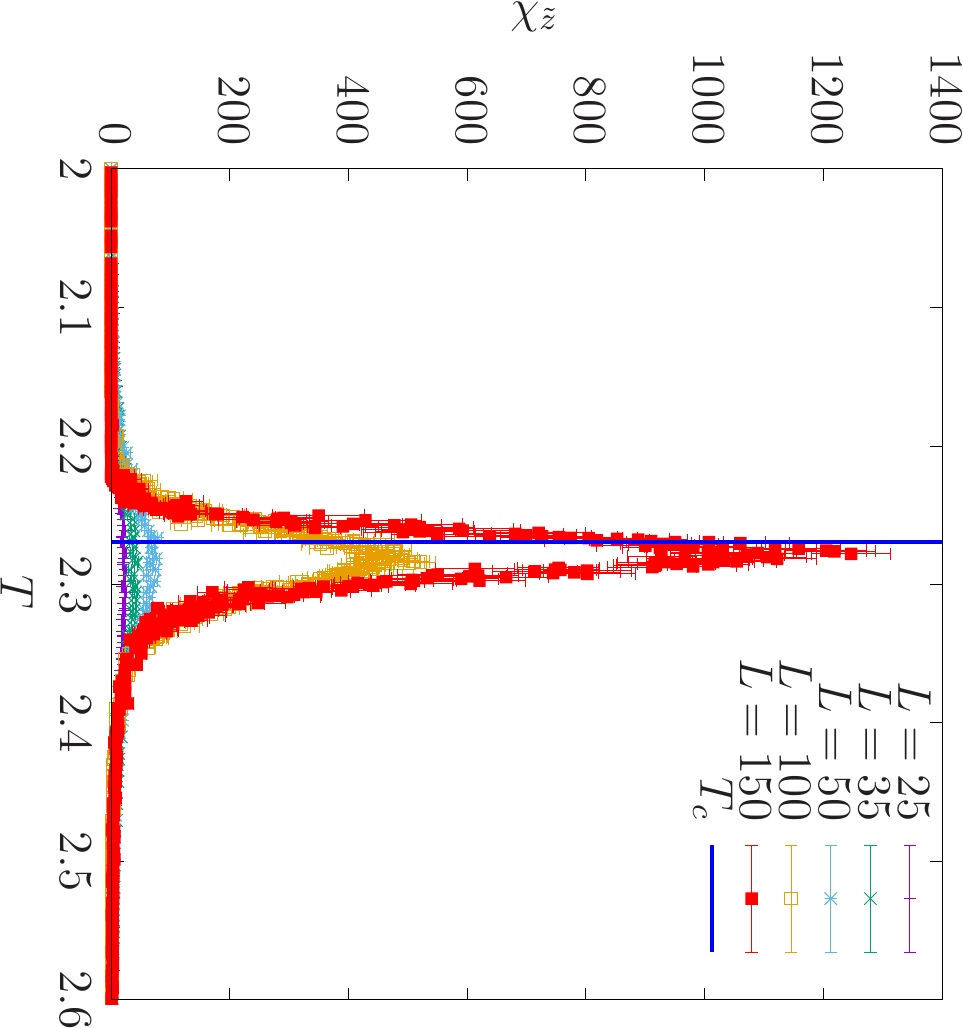}}
    \rotatebox{90}{\includegraphics[width=7.00cm]{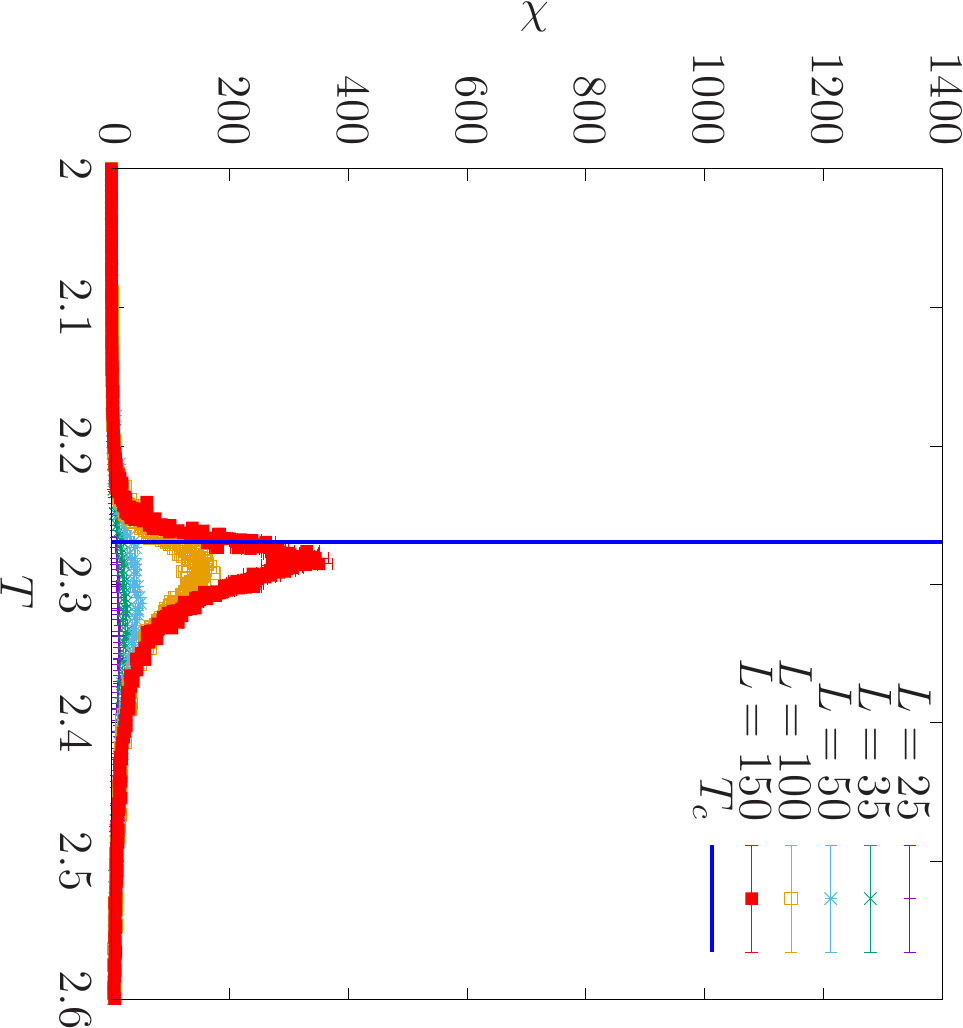}} 
\end{center}
\caption{The latent susceptibility (left) and  magnetic susceptibility (right) as a function of the temperature for five different lattice volumes. The blue vertical line denotes the analytically extracted value for the critical temperature (Eq.~\ref{eq:theoretically_extracted}). {\color{black}Each data point has been extracted as an ensemble average of the observable for fixed volume and temperature.}}
\label{fig:magnetic_susceptibility}
\end{figure*}

\subsection{The Latent Susceptibility and the Critical Temperature}
\label{subsec:criticaltemperature}
In the previous sections, we provided strong evidence that the latent {\color{black} variable}, resulting from the proposed autoencoder, demonstrates the underlying phase transition and that it can also be used as a rough estimate for the critical temperature $T_c$. Nevertheless, as the finite lattice size $L$ increases we need to make sure that $T_c(L)$ tends to the right limit, i.e. it convergences to the theoretically extracted value given in Eq.~\ref{eq:theoretically_extracted} as $L \to \infty$.

To investigate the convergence of $T_c(L)$, we first extract $T_c(L)$ for each different lattice size and then extrapolate to infinite $L$. $T_c(L)$ can be extracted by probing the peak of the latent susceptibility for each $L$. The latent susceptibility as a function of the temperature for the five different lattice sizes, is presented on the left-hand side of Fig.~\ref{fig:magnetic_susceptibility}. Unlike the magnetic susceptibility, presented on the right-hand side of Fig.~\ref{fig:magnetic_susceptibility}, the latent susceptibility is much sharper with peaks being closer to the known critical temperature $T_c$. This means that the critical temperature for each $L$ is influenced by less finite-size scaling effects.

Our {\color{black} temperature} grid is fine enough and enables an adequate extraction of the $T_c(L)$ from the coordinates of Fig.~\ref{fig:magnetic_susceptibility}. Hence, there is no need to use   multi-histogram reweighting~\cite{Giannetti:2018vif} techniques. The latent {\color{black} variable} behaves to a large extent as a step function, and thus, tends to $\propto \delta(T-T_c)$ as $L \to \infty$. In addition, the derivative of the latent susceptibility appears to be continuous. So we can also use a Gaussian fit to estimate the critical temperature.

\begin{figure}[!ht]
\vspace{1cm}
\begin{center}
    \rotatebox{0}{\includegraphics[width=8cm]{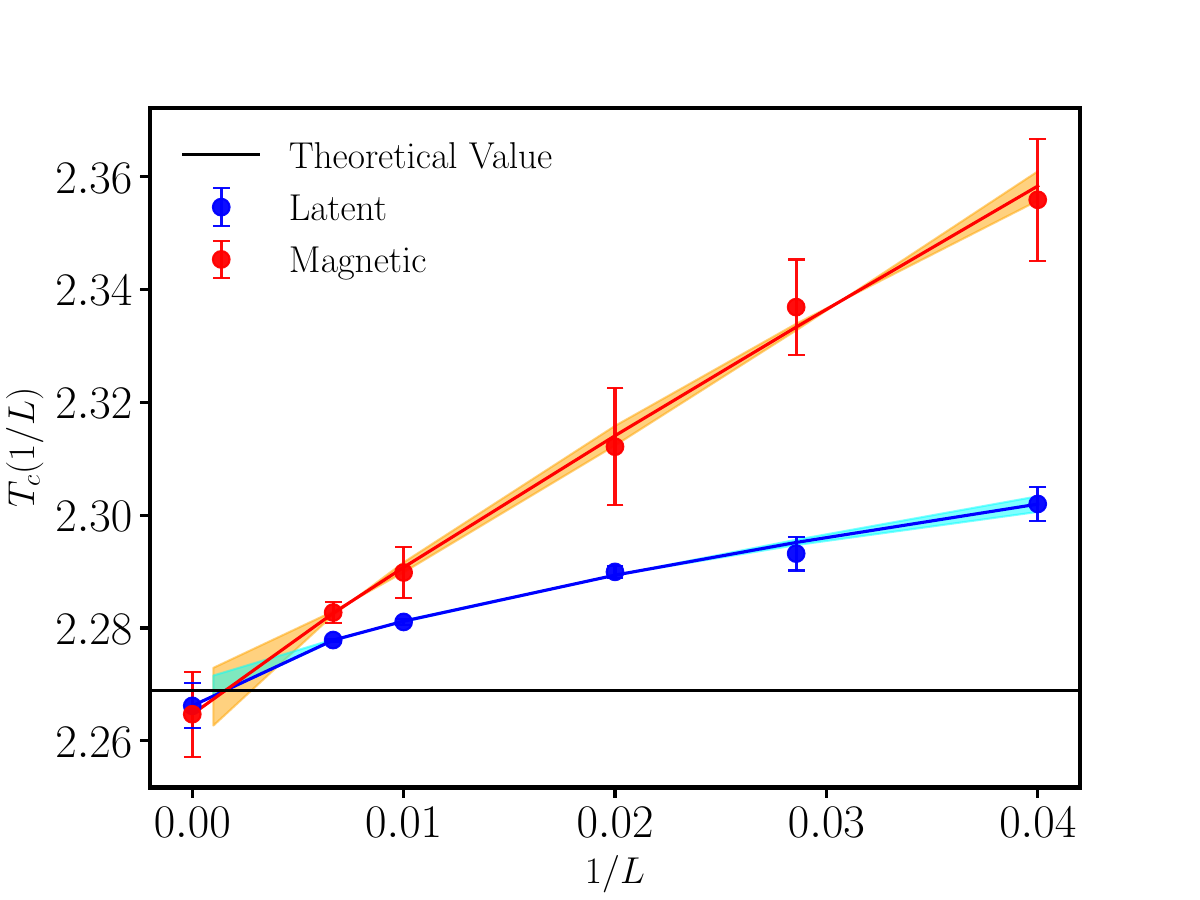}}
\end{center}
\caption{The critical temperature $T_c(L)$ extracted from fitting  the  magnetic (red) and  the latent (blue) susceptibilities as a function of $1/L$ to Eq.~\ref{eq:ansatz}. The error bands are estimated using the jackknife fit errors on the fit parameters.}
\label{fig:critical}
\end{figure}
In Fig.~\ref{fig:critical} we present $T_c(L)$ extracted from fitting the latent susceptibility and the magnetic susceptibility as a function of $1/L$. Results obtained using the latent susceptibility suffer less from finite-size scaling effects as compared to those when using the magnetic susceptibility. Adopting, the usual finite-size scaling behaviour

\begin{eqnarray}
T_c(L) - T_c(L=\infty) \propto L^{-1/\nu} \,,
\label{eq:ansatz}
\end{eqnarray}
\\
\noindent
we fit both susceptibilities to the ansatz $T_c(L) = T_c(L=\infty) + \alpha L^{-1/\nu}$.
Our findings are listed in Table~\ref{tab:fits-ML}.
\\
\begin{table}[h!]
        \begin{center}
	    \begin{tabular}{c|ccc}
		    \hline \hline
	    Susceptibility & $T_c(L=\infty)$ & $\nu$ & $\chi^2/{\rm dof}$ \\
		    \hline \hline
	    Magnetic &	$2.265(8)$ & $1.08(20)$ & $0.15$ \\
	    Latent   &	$2.266(4)$ & $1.60(14)$ & $0.41$ \\
		    \hline \hline
	    \end{tabular} \\
	    \caption{The results for $T_c(L=\infty)$ and $\nu$ extracted by fitting the magnetic as well as the latent susceptibilities to the ansatz $T_c(L) = T_c(L=\infty) + \alpha L^{-1/\nu}$.}
	    \label{tab:fits-ML}
	    \end{center}
\end{table}

As expected, fitting the data for $T_c(L)$ resulting from the magnetic susceptibility yields values of $T_c(L=\infty)$ and $\nu$ which are consistent with the analytically extracted values $T_c=2.269184$ and $\nu=1$. Turning now to the case of the latent {\color{black} variable}, it appears that the results of $T_c(L)$ when fitted with a form of the known scaling behaviour of Eq.~\ref{eq:ansatz}, yield a value for $T_c(L=\infty)$, which is in accordance with the theoretical expectation. One can observe that the $T_c$ curves in the infinite volume limit intersect with the theoretical value in the thermodynamic limit. This provides good evidence that the deep learning autoencoder does not only predict the phase regimes of the 2D-Ising model as well as give an estimate for the critical temperature but can also lead to a precise evaluation of the critical temperature.

\section{Results for the anti-ferromagnetic Ising model}
\label{sec:results_antiferromagnetic}

Having demonstrated the use of the autoencoder on the 2D-Ising ferromagnetic model, we turn now to the anti-ferromagnetic where we simply test the application of the network on the produced configurations. We investigate how the latent {\color{black} variable} $z_{i_{\rm conf}}$ behaves as a function of the temperature $T$ for each configuration. We generate $6000$ configurations, namely $200$ configurations for every single temperature. The configurations are for $30$ different values of temperatures within the range $T=1-4.5$. Once more, we make sure that we cover the whole range of temperatures between the two extreme cases of the anti-ferromagnetic Ising behaviour, the nearly "frozen" at $T \simeq 1$ and completely disordered $T \simeq 4.5$. We choose the temperature grid to be dense close to the theoretical, critical temperature. Our assumption is that since the ferromagnetic is connected to the anti-ferromagnetic via a bijective map between the spin fields, the autoencoder should be able to "notice" the phase transition. Hence, we tempted to check whether one can approximately detect the critical temperature using the latent variable on a small number of configurations.

\begin{figure*}[!h]
  \begin{center}
\includegraphics[width=17cm]{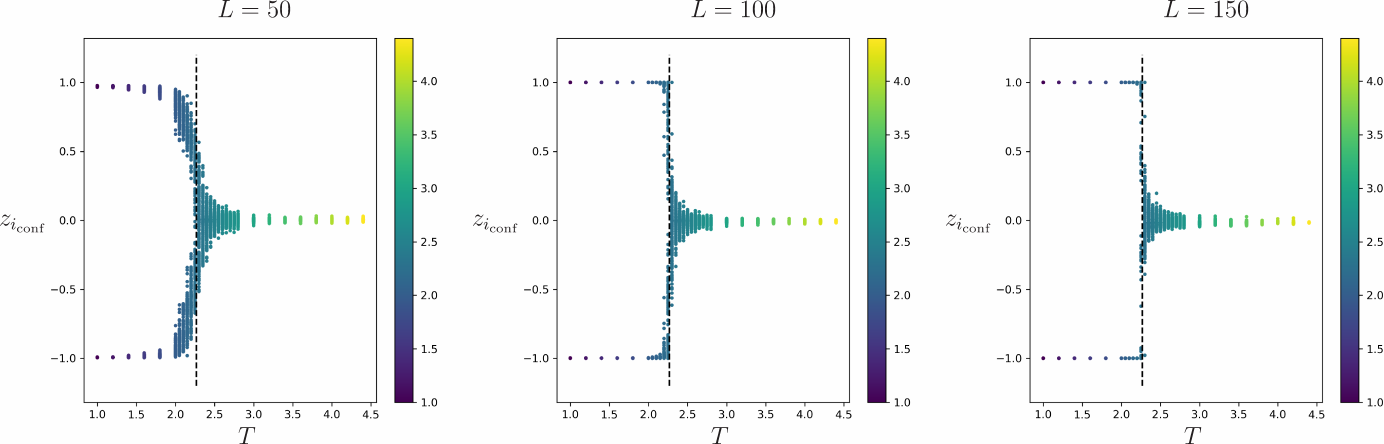}
  \end{center}
\caption{The latent {\color{black} variable} for each configuration as a function of the temperature for three different lattice volumes for the anti-ferromagnetic Ising model. {\color{black} These are scatter plots where no averaging was done for every single input data}. The dashed line represents the analytically extracted value of the critical temperature (Eq.~\ref{eq:theoretically_extracted}). The color on the gradient illustrator on the right denotes the temperature $T$.}
\label{fig:autoencoder_model_antiferromagnetic}
\end{figure*}

For the case of the anti-ferromagnetic Ising model we restrict the analysis to lattice volumes of $L=50$, $100$ and $150$. After training the autoencoder on configurations of the anti-ferromagnetic 2D-Ising model the reconstruction error was found to range from MSE of $0.68$, MSE of $0.69$ and MSE of $0.705$ for $L=50$, $100$ and $150$ respectively.  In Fig.~\ref{fig:autoencoder_model_antiferromagnetic} we demonstrate the latent {\color{black} variable} for each different configuration, as a function of the temperature $T$, for three different lattice sizes, $L=50$, $100$ and $150$. The behaviour of the latent {\color{black} variable} $z_{i_{\rm conf}}$ in these plots resembles the features of the latent {\color{black} variable} $z_{i_{\rm conf}}$ for the ferromagnetic 2D-Ising model presented in Fig.~\ref{fig:autoencoder_model}.

It is markedly observed upon representing the latent variable as a function of temperature that the autoencoder "notices" the two phases also for the case of the anti-ferromagnetic 2D-Ising model and provides a good approximation of the critical temperature. This reflects the spontaneously broken $\mathbb{Z}_2 \equiv \{-1,1\}$ symmetry group for the bijectively mapped sub-lattices as well as the spontaneously broken translation symmetry for the anti-ferromagnetic model. In Fig.~\ref{fig:autoencoder_model_antiferromagnetic} the plot for $L=150$ indicates that the transition occurs right at the critical temperature. One can fit the points of the latent variable which behave linearly to a constant as a function of the temperature and can restrict that the collapse of the two states located at 1 and -1 occurs at $T=2.288(21)$. This is in good agreement with the theoretical prediction.

\begin{figure*}[!ht]
  \begin{center}
\rotatebox{0}{\includegraphics[width=14cm]{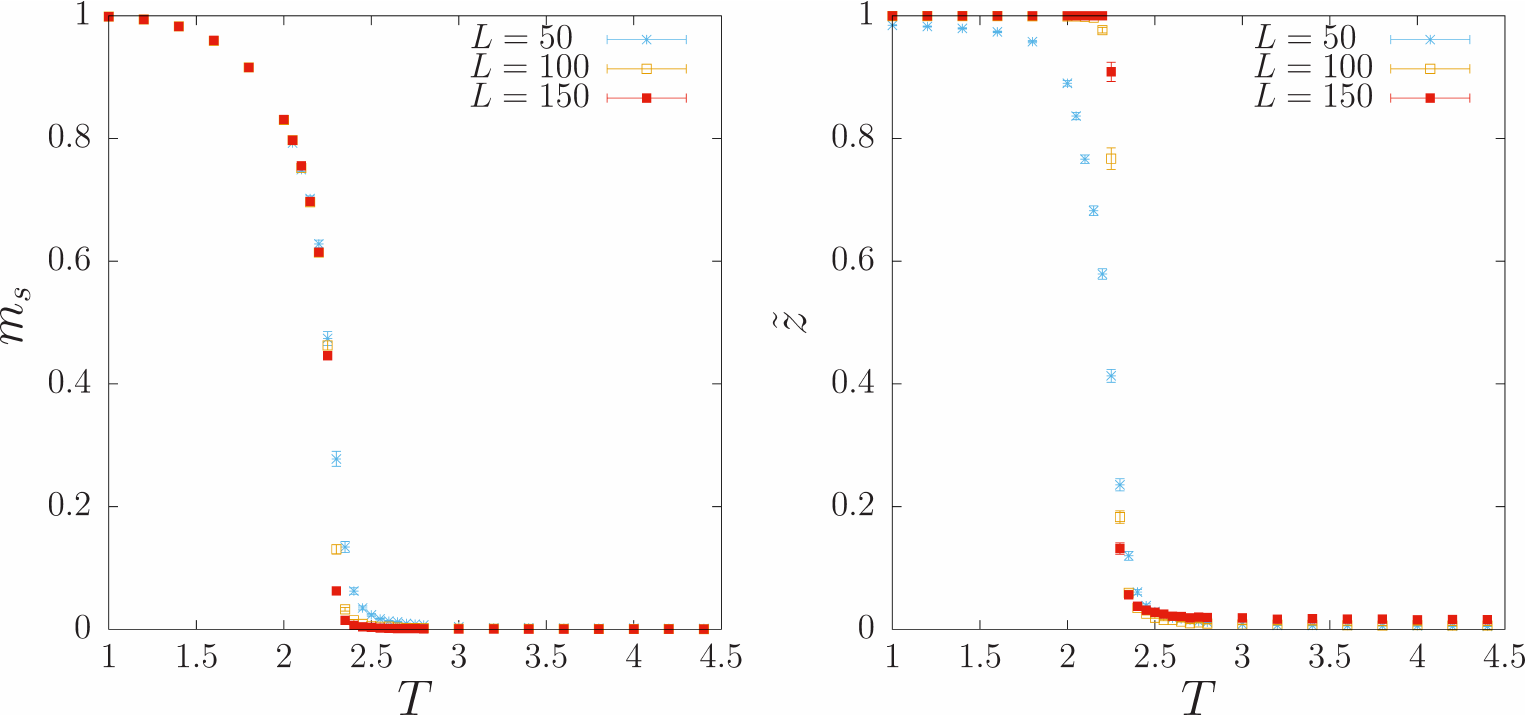}}
  \end{center}
\caption{\color{black} The average absolute staggered magnetization (left) compared to the average latent {\color{black} variable} (right) as a function of the temperature for the anti-ferromagnetic Ising model and three different lattice volumes. The data presented on the right plot stem from testing data. {\color{black}Each data point has been extracted as an ensemble average of the observable for fixed volume and temperature.}}
\label{fig:average_latent_antiferromagnetic}
\end{figure*}

On the left-hand-side of Fig.~\ref{fig:average_latent_antiferromagnetic} we present the staggered magnetization as a function of the temperature while on the right-hand side we provide the absolute latent {\color{black} variable} for the anti-ferromagnetic model. As for the case of the ferromagnetic model, the absolute latent {\color{black} variable} looks similar to the staggered magnetization, albeit becoming steeper as the lattice size increases. This demonstrates that the absolute latent variable and as a consequence the latent susceptibility can enable an extraction of the critical temperature with smaller scaling effects compare to the staggered magnetization and susceptibility respectively. Further analysis of the anti-ferromagnetic Ising model will be presented in~\cite{Alexandrou:inpreparation}.  

\section{Conclusions and Outlook}
\label{sec:conclusions}
In this work, we apply a deep learning auto-encoder on configurations produced for the 2D (anti)ferromagnetic Ising model for performing classification in an unsupervised manner. Hence, with no prior knowledge on the system, we demonstrate that we can predict the phase structure of this system qualitatively as well as quantitatively by determining both phase regions and the critical temperature. 

For the ferromagnetic model, at low temperatures, by making use of the latent {\color{black} variable} per configuration, the autoencoder predicts two states reflecting to the broken $\mathbb{Z}_2$ symmetry. As the temperature increases, these two states appear to collapse at one state, located around zero, and the underlying symmetry is restored. This behaviour becomes more distinct as the volume of the lattice increases, and the point where the two states collapse is getting more and more local; this corresponds to the critical point of the phase transition.

One can define the average absolute latent {\color{black} variable} ${\tilde z}$ that displays partially the characteristics of an order parameter; namely, it can identify the phase but cannot capture the order of the phase. Although it resembles the behaviour of the magnetization, it becomes steeper as the size of the volume increases, tending to a step function. The second moment of the absolute latent {\color{black} variable} defines a susceptibility, named latent susceptibility, the peak of which can determine the critical temperature $T_c(L)$. By extrapolating the values of $T_c(L)$  to $L \to \infty$ for the sequence of lattice sizes $L=25, \ 35, \ 50, \ 100, \ 150$, we obtain for $T_c(L = \infty)= 2.266(4)$ in  agreement with the exact value of $T_c=2.26918$ calculated analytically. This suggests that the proposed deep learning (fully-connected) autoencoder can identify, in an unsupervised manner, the phase structure of the 2D ferromagnetic Ising model but can also lead to a precise extraction of the critical temperature at the limit of the infinite volume. As shown in Fig.~\ref{fig:critical} the values of $T_c(L)$ suffer with less finite size effects compared to those usually extracted by using the peak of the magnetic susceptibility, and one would thus expect that the autoencoder could give a more precise prediction for $T_c$. Of course to test this hypothesis we need to extract $T_c(L)$ for larger volumes, for instance up to $L=1024$ similarly to Ref.~\cite{Giannetti:2018vif}, and obtain the extrapolated value of $T_c(L=\infty)$. This requires the usage of a different autoencoder with more layers since memory limitations make the current autoencoder insufficient to work. This is a future extension of this work.

Applying the deep learning auto-encoder on configurations produced for the 2D anti-ferromagnetic Ising model, we observe that the results resemble to an adequate extent those for the ferromagnetic model. Namely, by using the latent {\color{black} variable} per configuration, the autoencoder predicts two states reflecting to the broken $\mathbb{Z}_2$ symmetry of the two bijectively connected sub-lattices as well as the broken translation symmetry. Once more, this behaviour becomes more distinct as we increase the lattice size with the collapse of the two states becoming steeper. This special point corresponds to the critical point of the phase transition. In the same manner as for the ferromagnetic case, we can make use of the average absolute latent variable which behaves similarly to the average stout magnetization, albeit becoming steeper with the lattice size.  

{\color{black}This work provides a good indication that, with the right choice of parameters, deep learning autoencoders can be used as tools to define new quantities which are affected less by finite scaling effects and lead to more precise evaluation of observables related to the phase structure of statistical models. This could be proven beneficial for theories in which the production of thermalized uncorrelated configurations requires a large number of computational resources.}

There are other several related directions in which this work can be extended. Since our proposed autoencoder has been tested only on the 2D-Ising model, it would be important to investigate its generalization to other physical systems with non-trivial phase structure. An important question, which could be answered is whether this neural network is capable of identifying the phases for cases in which an order parameter is either not known or not existing; such an example is the Hubbard model~\cite{1963RSPSA.276..238H} describing the transition between conducting and insulating systems. Another relevant question is how {\color{black} this particular} autoencoder behaves in cases where the phase transition is of a different order, or an infinite order such as in the 2D $XY$\footnote{Autoencoders have already been applied to the 2D XY model~\cite{Cristoforetti:2017naf}, albeit as a generative procedure in order to support the Monte-Carlo simulation of the system and production of configurations.} spin model where the relevant phase transition is the Kosterlitz-Thouless which is of infinite order~\cite{Kosterlitz:1973xp}. Finally, our future plans involve the testing of the autoencoder as a tool for the unsupervised extraction of the phase structure of physical systems with continuous symmetries. These involve quantum field theories formulated on the lattice such as the 3D $\phi^4$ with $O(2)$ symmetry~\cite{Arnold:2001ir} where the phase transition is of second order and belongs to the same universality class as the 2D-Ising model, the 3D $U(1)$ gauge theory~\cite{Athenodorou:2018sab} for which the phase transition is of infinite order and belongs to the same universality class as the 2D $XY$ model, as well as the 3D $SU(N)$ gauge theory~\cite{Liddle:2008kk} which has a second-order phase transition for $N \leq 3$, a weakly first order for $N=4$ and the first order for $N \geq 5$.

\begin{acknowledgements}
We would like to thank Giannis Koutsou, Nikos Savva, Spyros Sotiriadis, Mike Teper and Savvas Zafeiropoulos for fruitful discussions. We would also like to express our gratitude to Barak Bringoltz, Biagio Lucini and Davide Vadacchino for performing a critical reading of the manuscript. AA was financially supported by the Horizon 2020 of the European Commission research and innovation programmes VI-SEEM under grant agreement No 675121, OpenSESAME under grant agreement No 730943 and Tips in SCQFT under grant agreement No 791122.  CC also acknowledges support from the OpenSESAME project. SP is supported by the Horizon 2020 of the European Commission research and innovation programmes HPC-LEAP under the Marie Sk\l odowska Curie grant agreement No. 642069 and PRACE-5IP under grant agreement No 730913. {\color{black}The authors express their gratitude to the anonymous reviewers for their helpful and constructive comments.}
\end{acknowledgements}

\end{document}